\documentclass[aps,preprintnumbers,showpacs,nofootinbib,superscriptaddress,floatfix,notitlepage]{revtex4-1}

\pdfoutput=1 

\usepackage{graphicx} \usepackage{amssymb} \usepackage{amsmath}
\usepackage{bm} \usepackage{slashed} \usepackage{datetime}
\usepackage{mciteplus} \usepackage{multirow} \usepackage{color}
\usepackage[utf8]{inputenc} \usepackage[normalem]{ulem}

\newcommand{\xbj}{x_B}
\newcommand{\zh}{z_h}
\newcommand{\de}{d}     

\newcommand{\xb}{\hat{x}} 
\newcommand{\zd}{\hat{z}}

\newcommand{\T}{\perp} 
 
\newcommand{\bT}{\xi_T}
 
\newcommand{\ms}{\mskip 1.5mu}

\usepackage[normalem]{ulem} 

\newcommand{\new}[1]{{\color[rgb]{0,0,1}{#1}}} 

\begin{document}
\allowdisplaybreaks[2]

\title{ Azimuthal asymmetries in unpolarized SIDIS and Drell-Yan
  processes: a case study towards TMD factorization at subleading
  twist}

\author{Alessandro Bacchetta} \email{alessandro.bacchetta@unipv.it}
\affiliation{Dipartimento di Fisica, Universit\`a di Pavia, via Bassi
  6, I-27100 Pavia}  \affiliation{INFN Sezione di Pavia, via Bassi 6,
  I-27100 Pavia, Italy}

\author{Giuseppe Bozzi} \email{giuseppe.bozzi@unipv.it}
\affiliation{Dipartimento di Fisica, Universit\`a di Pavia, via Bassi
  6, I-27100 Pavia}  \affiliation{INFN Sezione di Pavia, via Bassi 6,
  I-27100 Pavia, Italy}

\author{Miguel G.~Echevarria} \email{mgechevarria@pv.infn.it}
\affiliation{INFN Sezione di Pavia, via Bassi 6, I-27100 Pavia, Italy}

\author{Cristian Pisano} \email{cristian.pisano@ca.infn.it}
\affiliation{Dipartimento di Fisica, Universit\`a di Cagliari,
  Cittadella Universitaria, I-09042 Monserrato (CA), Italy }
\affiliation{INFN Sezione di Pavia, Cittadella Universitaria, I-09042
  Monserrato (CA), Italy}

\author{Alexey Prokudin} \email{prokudin@jlab.org}
\affiliation{Science Division, Penn State University Berks, Reading,
  Pennsylvania 19610, USA}  \affiliation{Jefferson Lab, 12000
  Jefferson Avenue, Newport News, Virginia 23606, USA}

\author{Marco Radici} \email{marco.radici@pv.infn.it}
\affiliation{INFN Sezione di Pavia, via Bassi 6, I-27100 Pavia, Italy}

\begin{abstract}
  We consider the azimuthal distribution of
  the final observed hadron in semi-inclusive deep-inelastic scattering
  and the lepton pair in the Drell-Yan process. In particular, we focus on the  
  $\cos \phi$ modulation of the unpolarized cross section and on its
  dependence upon transverse momentum. At low transverse momentum, 
  for these observables we propose a factorized expression based on tree-level 
  approach and conjecture that the same formula is valid in
  transverse-momentum dependent (TMD) factorization when written in terms of
  subtracted TMD parton distributions. Our formula correctly
  matches with the collinear factorization results at high transverse
  momentum, solves a long-standing problem and is a 
  necessary step towards the extension of the TMD factorization
  theorems up to the subleading twist.
\end{abstract}

\preprint{JLAB-THY-19-2963}

\pacs{13.60.Le, 13.87.Fh,14.20.Dh}

\maketitle

\section{Introduction}
\label{s:intro}

The inclusive production of a system of one or more particles with a specific transverse
momentum in lepton-hadron or hadron-hadron collisions is in general
characterized by three different scales:  the nonperturbative QCD
scale $\Lambda_{\rm QCD}$, the hard scale of the process $Q$, 
and the magnitude of the system's transverse momentum $q_T$. In both processes under study, namely
semi-inclusive deep inelastic scattering (SIDIS), $\ell p \to \ell^\prime h X$, and production of Drell--Yan lepton 
pairs (DY), $p p \to \ell^+ \ell^- X$, the hard scale $Q$ is given by
the virtuality of the gauge boson exchanged in the reaction.  

Depending on the value of the transverse momentum $q_T$, two different
frameworks are adopted for the description of these processes. 
At  {\em high $q_T$}, namely  $q_T \gg \Lambda_{\rm QCD}$,
the transverse momentum in the final state is generated by
the perturbative radiation and the cross section can be expressed in terms
of collinear ({\it i.e.}, integrated over transverse momentum) parton
distributions (PDFs) and fragmentation functions (FFs). Conversely,
at {\em low $q_T$}, $q_T \ll Q$, transverse-momentum-dependent (TMD)
factorization~\cite{Collins:2011zzd,GarciaEchevarria:2011rb,Echevarria:2012js}
can be applied and TMD PDFs and FFs (or TMDs for short)  dependent on the transverse momentum,  are used in the factorized expression. 
In principle,  in the intermediate region $\Lambda_{\rm QCD} \ll  q_T \ll Q$ both frameworks can be applied.
In case they describe the same mechanism (characterized  by the same power behavior), in this region they have to
match. If, on the other hand, the two results describe competing mechanisms, they 
should be considered independently and added together (see Ref.~\cite{Bacchetta:2008xw}).  

For the $q_T-$differential unpolarized cross section integrated over the azimuthal angle of the final particle,
the matching of the TMD and the collinear factorization descriptions in the intermediate $q_T$ region has been shown in, e.g.,
Refs.~\cite{Collins:1984kg,Catani:2000vq}. 
These results underpin all phenomenological studies of TMDs, even though some modifications of
the original procedure are often needed~\cite{Gamberg:2017jha,Echevarria:2018qyi}.  

Apart from the azimuthally independent cross sections, where unpolarized TMDs are important, various TMDs
are involved in generating azimuthal modulations of unpolarized cross sections. Both in SIDIS and DY, neglecting
parity-violating interactions, four structure functions are needed to
parametrize the cross section that depends on the azimuthal angle and transverse momentum. 
Two of them are related to what is usually referred to as $\cos\phi$ and $\cos 2\phi$ modulations, 
where $\phi$ is the azimuthal angle between the leptonic and hadronic planes in a specific frame. 

In this paper, we study $\cos \phi$ modulations:
they involve twist-3 TMD PDFs and FFs and are suppressed by a factor
$1/Q$ with respect to the leading (twist-2) terms.  These modulations
have been already studied,
 with somewhat contradictory results, in
Refs.~\cite{Boer:2006eq,Berger:2007jw,Chen:2016hgw} for
DY and in Ref.~\cite{Bacchetta:2008xw} for SIDIS. 
We remark that no factorization proof for TMD
observables at twist-3 is available, although steps in this
  direction have recently been taken (see
  e.g. Refs.~\cite{Feige:2017zci,Balitsky:2017flc,Balitsky:2017gis,Ebert:2018lzn,Ebert:2018gsn,Moult:2019mog}).
A yet unsolved problem is 
matching between the TMD and collinear descriptions in the intermediate $q_T$ region, which might
point to the conclusion that observables related to twist-3 TMDs cannot be properly factorized. 

We propose here a solution to the problem of matching the TMD and
collinear formulae. We suggest a TMD factorized formula modified with respect to the one used in Ref.~\cite{Bacchetta:2008xw}. 
We provide arguments in favor of this formulation and show that it leads to an agreement between the TMD
and collinear results in SIDIS and in DY, in different reference frames.
The leading logarithmic (LL) terms match also in the so-called Wandzura--Wilczek approximation~\cite{Wandzura:1977qf}.
Our results can be generalized to other observables involving twist-3 TMDs and they
represent a necessary step in the direction of establishing TMD factorization at twist-3 level.

\section{\new{Azimuthal} $\cos\phi$ asymmetry in SIDIS}
\label{s:SIDIS}

We start from a detailed discussion of the semi-inclusive deep inelastic scattering process
\begin{equation}
  \label{sidis}
\ell(l) + N(P) \to \ell(l') + h(P_h) + X (P_X),
\end{equation}
where $\ell$($\ell'$) is the incoming (outgoing) lepton with momentum
$l(l')$, $N$ is the nucleon with mass $M$ and momentum $P$, and $h$ is
the detected hadron with mass $M_h$ and momentum $P_h$.  We adopt the
standard SIDIS variables
\begin{align}
\xbj &= \frac{Q^2}{2\,P\cdot q} \,, & y &= \frac{P \cdot q}{P \cdot l}
\,, & \zh &= \frac{P \cdot P_h}{P\cdot q} \,,
\end{align} 
where $q=l-l^\prime$ and $Q^2=-q^2$. 

In the one-photon-exchange approximation, the differential cross section can be written as (see, e.g., Ref.~\cite{Barone:2003fy}) 
\begin{align}
\frac{d^5\sigma}{d\xbj\,dy\,d\zh\,d^2 \bm{P}_{hT}} &= \frac{\pi
  \alpha^2}{2Q^4} y L_{\mu\nu} W^{\mu\nu} \,,
\end{align}
with $\alpha$ being the fine structure constant and $L_{\mu\nu}$ and $W^{\mu\nu}$ the leptonic and hadronic tensors, respectively. 

The process is usually studied in a frame where $\bm{P}$ and $\bm{q}$
are collinear and taken to be along the $z$-axis, with the azimuthal
angle $\phi_h$ of the final hadron defined w.r.t.~the lepton plane
according to the so-called Trento
conventions~\cite{Bacchetta:2004jz}. We denote  by $P_{hT}$ the
component of $P_{h}$ transverse to the momenta $P$ and
$q$. Alternatively, one can choose $\bm{P}$ and $\bm{P}_{h}$ as the
longitudinal directions: in this case, the photon will carry a
transverse momentum $q_T$ related to $P_{hT}$ by the relation 
\begin{equation}
q_T^\mu = -\frac{1}{z}\,P^\mu_{hT} - 2 \rho^2 x P^\mu\,,
\label{eq:qT}
\end{equation}
with $\rho^2 = q_T^2/Q^2$ and $q_T^2 = - q_T^\mu \,q_{T\mu} \equiv
\bm{q}_T^2$.~\footnote{If there is no ambiguity, in the
    following we will use the notation $a_T$ for the modulus of the
    spatial vector $|\bm{a}_T|$.}

Moreover, we limit ourselves to a kinematic region where 
$Q^2 \gg \Lambda_{\rm QCD}^2\approx M^2$ at
fixed values of $x$, $y$, $z$, and neglect corrections of order
$M^2/Q^2$.  The cross section can be parametrized in terms of four
structure functions that depend on $x$, $z$, $Q^2$ and
$P^2_{hT}$~\cite{Bacchetta:2006tn}, 
\begin{align}
\frac{d\sigma}{dx \, dy \,dz \, d\phi_h\, d P_{hT}^2}  = &
\frac{\pi\alpha^2}{x \ms Q^2}\, \frac{y}{1-\varepsilon} 
\biggl\{ F_{UU ,T}+  \varepsilon\ms F_{UU ,L} + \sqrt{2\,\varepsilon
  (1+\varepsilon)}\,\cos\phi_h\, F_{UU}^{\cos\phi_h} + \varepsilon
\cos 2\phi_h\,  F_{UU}^{\cos 2\phi_h} \bigg \} \,,
\label{eq:cs-SIDIS-F}
\end{align}
where $\varepsilon$ is the
ratio of the longitudinal and transverse photon fluxes,
\begin{equation}
\varepsilon = \frac{1-y}{1-y+ y^2/2}~.
\end{equation} 
The first and
second subscripts of the structure functions refer to the polarization of the initial lepton and
proton, respectively, while the third one specifies the polarization
of the virtual photon exchanged in the reaction.

We focus here on the structure functions $F_{UU ,T}$ and
$F_{UU}^{\cos\phi_h}$. At tree level, they can be written in terms of
TMDs in the following way
\begin{align} 
\label{F_UUT0}
F_{UU ,T}\bigl(x,z,P_{hT}^2,Q^2\bigr) & 
= \sum_a
e_a^2 x \mathcal{B}_0 \Bigl[ \widehat{f}_1^a \, \widehat{D}_1^a \Bigr]
 +{\cal O}\biggl(\frac{P_{hT}^{2}}{Q^{2}}\biggr)\; ,\\
\begin{split} 
F_{UU}^{\cos\phi_h}\bigl(x,z,P_{hT}^2,Q^2\bigr) & 
= \sum_a e_a^2 x
\frac{2M M_h}{Q}\,
\\
&\quad \mathcal{B}_1\biggl[ x  \widehat{h}^a \,
  \widehat{H}_{1}^{\perp a (1)}  + \frac{M_h}{M}\,\widehat{f}_1^a \,
  \frac{\smash{\widehat{\tilde{D}}}^{\perp a (1)}}{z}
  -
  \frac{M}{M_h} x \widehat{f}^{\perp a (1)} \, \widehat{D}_1^a -
  \widehat{h}_{1}^{\perp a (1)} \,  \frac{\widehat{\tilde{H}}^a}{z}
  \biggr]
  +{\cal O}\biggl(\frac{P_{hT}^{2}}{Q^{2}}\biggr)\; ,
   \label{F_UUcosphi0}
\end{split} 
\end{align}  
where 
\begin{equation}
  \mathcal{B}_n \big[ \widehat{f} \, \widehat{D} \big] = 2\pi
  \int_0^{\infty} {d \bT} \, \bT^{n+1} J_n\biggl( \frac{\bT
  P_{hT}}{z} \biggr) \,  \widehat{f}^a \big(x, \bT^2; Q^2\big) \,
\widehat{D}^a \big(z, \bT^2; Q^2\big) \,,
\end{equation} 
the $e_a$ is the electric charge of a parton with flavor $a$ in units
of the proton charge, and the Fourier transform of a generic TMD PDF, $f(x,k_\T^2;Q^2)$,
has been defined as
\begin{align} 
\widehat{f}\big(x, \bT^2; Q^2\big) &\equiv  \frac{1}{2\pi}\,\int
d^2\bm{k}_\T e^{i {\bm \xi}_T  \cdot \bm k_\perp}  f \big(x,
k_\T^2;Q^2\big) = \int_0^{\infty} d k_\T  k_\T \, J_0\big(\bT k_\T
\big) \,  f \big(x, k_\T^2; Q^2)\,
. 
\end{align}
Furthermore, the $ \bT^2$-derivatives of the TMDs are given
by~\cite{Boer:2011xd}
\begin{align}
\widehat{f}^{(n)}(x, \bT^2; Q^2) &= n !\bigg(-\frac{2}{M^2}
\frac{\partial}{\partial \bT^2}\bigg)^n \widehat{f} (x, \bT^2; Q^2)  =
\frac{n!}{M^{2 n}} \int_0^{\infty} d k_\T k_\T \, \biggl( \frac{
  k_\T}{\bT} \biggr)^n \,   J_n\big(\bT k_\T \big) \, f \big(x,
k_\T^2; Q^2 \big)\,
.
\label{eq:f-der}
\end{align} 
For a generic TMD FF, $D(z,P_{\T}^2/z^{2};Q^2)$, the above formulas are identical, 
but with $k_\T$ replaced by $P_{\T}/z$.

For the structure function $F_{UU ,T}$, and in general for twist-2
terms in the hadronic tensor, a general factorization proof can be given, 
see for example Ref.~\cite{Collins:2011zzd}.
Soft gluon radiation to all orders is absorbed into an exponential Sudakov form factor which is 
partitioned between TMD PDFs and FFs, while all the remaining perturbative corrections are incorporated 
in the so-called hard factor $\mathcal{H}$. The final formula resembles very much the tree-level result: 
\begin{align} 
\label{F_UUT}
F_{UU,T}(x,z, P_{hT}^2, Q^2)  &= \mathcal{H}_{\mathrm{SIDIS}} (Q^2,\mu^2) \, \sum_a
e_a^2 x \, \mathcal{B}_0 \Bigl[ \widehat{f}_1^a (x,\xi^2_T;\mu^2,\nu^2)\, \widehat{D}_1^a (z,\xi^2_T;\mu^2,\nu^2) \Bigr]
\,  +{\cal O}\biggl(\frac{P_{hT}^{2}}{Q^{2}}\biggr)\; ,
\end{align} 
where the scales $\mu$ and $\nu$ arise as a consequence of regulating the ultraviolet and rapidity divergences of TMDs. 
These scales can be both set equal to $Q$ in order to minimize logarithmic
corrections. 
In the following, we will refer to these properly defined TMD functions as {\em subtracted} TMDs.

For what concerns the structure function $F_{UU}^{\cos\phi_h}$ (and in
general twist-3 terms in the hadronic tensor), we {\em conjecture}
that the correct formula can be constructed in the same way as for
twist-2 terms, i.e., starting from the tree-level formula, adding the 
hard scattering function $\mathcal{H}$ and  replacing TMDs with subtracted ones.
We found no obvious way to prove this conjecture. However, what gives us confidence in its
validity is that the resulting formula correctly matches the perturbative calculation at
high transverse momentum, both in SIDIS and DY, thus
fixing the mismatch observed in Sec.~8.3 of Ref.~\cite{Bacchetta:2008xw}.

Our starting formula is therefore
\begin{equation}
  \begin{split} 
F_{UU}^{\cos\phi_h}\bigl(x,z,P_{hT}^2,Q^2\bigr)  
& = \frac{2M M_h}{Q}\,\mathcal{H}'_{\mathrm{SIDIS}} \, \sum_a
e_a^2 x \,
\\
&\quad \times
\mathcal{B}_1\Biggl[   x  \widehat{h}^a \, \widehat{H}_{1}^{\perp a (1)}
  + \frac{M_h}{M}\,\widehat{f}_1^a \,
  \frac{\smash{\widehat{\tilde{D}}}^{\perp a (1)}}{z}
   - \frac{M}{M_h} x \widehat{f}^{\perp a (1)} \, \widehat{D}_1^a -
  \widehat{h}_{1}^{\perp a (1)} \, \frac{\widehat{\tilde{H}}^a}{z} \Biggr]
+{\cal O}\biggl(\frac{P_{hT}^{2}}{Q^{2}}\biggr)
\,.
   \label{e:F_UUcosphi}
\end{split} 
\end{equation}
This structure function has been analyzed  by
Cahn~\cite{Cahn:1978se,Cahn:1989yf}, who for the first time pointed
out  the presence of perturbative and nonperturbative
contributions. Measurements of this azimuthal modulation are available
in Refs.~\cite{Arneodo:1986cf,Adams:1993hs,Breitweg:2000qh,Chekanov:2006gt,Airapetian:2012yg,Adolph:2014pwc,Moretti:2019lkw}. 
Phenomenological analyses that took into account various contributions
separately have been reported in Ref.~\cite{Anselmino:2005nn,Anselmino:2006rv,Barone:2015ksa}.  
 
\subsection{From high to intermediate transverse-momentum}

In the high transverse-momentum region ($q_T \gg \Lambda_{\rm QCD} $)
structure functions can be expressed,  using
collinear factorization, in terms of convolutions of hard scattering
coefficients with the usual collinear distribution
and fragmentation functions, respectively denoted by $f_1$ and
$D_1$~\cite{Mendez:1978zx,Bacchetta:2008xw}.  To the first order in the strong coupling $\alpha_s$,
this result can be further approximated in the intermediate
transverse-momentum region ($\Lambda_{\rm QCD} \ll q_T \ll Q$)
as~\cite{Meng:1995yn,Mendez:1978zx,Bacchetta:2008xw}
\begin{align}
  \begin{split} 
  \label{e:high_FUU}
F_{UU, T} 
&= \frac{1}{q_T^2}\, \frac{\alpha_s}{2\pi^2 z^2} {\Bigg\{ }
\sum_a x  e_a^2\,
\biggl[f_1^a(x, {Q^2})\,D_1^a(z, {Q^2})\,L\biggl( \frac{Q^2}{q_T^2} \biggr)
+ f_1^a(x, {Q^2})\, \bigl( D_1^a \otimes P_{qq}
+ D_1^g \otimes P_{gq} \bigr)(z, {Q^2})
\\
& \quad
+ \bigl( P_{qq} \otimes f_1^a 
+ P_{qg} \otimes f_1^g \bigr)(x, {Q^2})\, D_1^a(z, {Q^2})
\biggr]
+{\cal O}\biggl(\frac{\Lambda_{\rm QCD}}{q_T}\biggr)
+{\cal O}\biggl(\frac{q_T}{Q}\biggr) {\Bigg\} }\,,
\end{split} 
\\
\begin{split} 
\label{e:high_FUUcosphi}
F_{UU}^{\cos\phi_h} &= - \frac{1}{Q q_T}\, \frac{\alpha_s}{2\pi^2 z^2} {\Bigg\{ }
\sum_a x e_a^2\, \biggl[f_1^a(x; Q^2)\,D_1^a(z; Q^2)\,L\biggl(
  \frac{Q^2}{q_T^2} \biggr) + \,\sum_{i=a,g} \biggl(f_1^a(x; Q^2) (D_1^i \otimes
  P_{ia}')(z; Q^2)
  \\ & \quad
  + (P_{ai}' \otimes f_{1}^{i})(x; Q^2)\, D_1^a(z; Q^2)\biggr)\biggr]
+{\cal O}\biggl(\frac{\Lambda_{\rm QCD}}{q_T}\biggr)
+{\cal O}\biggl(\frac{q_T}{Q}\biggr) {\Bigg\} }\,,
\end{split}
\end{align}
where the factor $L$ is defined as 
\begin{equation}
L\biggl( \frac{Q^2}{q_T^2}\biggr) = 2 C_F \ln \frac{Q^2}{q_T^2} - 3
C_F \,.
\label{e:sudakovleading}
\end{equation} 
The convolutions are defined as 
\begin{align}
  \label{con-def}
\bigl(C \otimes f \bigr)(x; Q^2) &= \int_{x}^1 \frac{\de\xb}{\xb}\;
C(\xb; Q^2)\, f\Bigl(\frac{x}{\xb}; Q^2\Bigr) \,,
& \bigl(D \otimes C \bigr)(z; Q^2) &= \int_{z}^1 \frac{\de\zd}{\zd}\;
D\Bigl(\frac{z}{\zd}; Q^2\Bigr)\, C(\zd; Q^2) \,,
\end{align}
and the splitting functions are given by
\begin{align} 
  \label{splitting-fcts}
P_{qq}(\xb) &= C_F \biggl[\frac{1+\xb^2}{(1-\xb)_{+}}
               + \frac{3}{2}\,\delta(1-\xb) \biggr] \,,
&
P_{qg}(\xb) &= T_R\, \bigl[\xb^2+(1-\xb)^2\bigr] \,,
&
P_{gq}(\xb) &=C_F\,\frac{1+(1-\xb)^2}{\xb} \,,
\\
P_{qq}'(\xb) &= C_F \biggl[\frac{2\xb^2}{(1-\xb)_{+}} +
  \frac{3}{2}\,\delta(1-\xb) \biggr] \,,
&
P_{qg}'(\xb) &= 2T_R\,
\xb\,(2\xb-1) \,,
&
P_{gq}'(\xb) &= -2 C_F\,(1-\xb)\, , \phantom{\frac{\xb^2}{\xb}}
\end{align}
where $C_F = (N_c^2-1)/2 N_c$, $N_c=3$ being the number of colors,
$T_R = 1/2,$ and the plus-distribution is defined by 
 \begin{equation}
  \label{plus-def}
\int_z^1 \de y \, \frac{G(y)}{(1-y)_{+}}  = \int_z^1 \de y \,
\frac{G(y)-G(1)}{1-y}  - G(1)\ms \ln\frac{1}{1-z} ~.
\end{equation} 

\subsection{From low to intermediate transverse-momentum}

We use Eq.~\eqref{e:F_UUcosphi} as our starting point.
The distribution functions $f^\perp$, $h$ and the fragmentation
functions $\tilde D^\perp$, $\tilde H$ 
are twist-3 TMDs.  
The QCD equations of motion (EOM) lead to the useful relations \cite{Bacchetta:2006tn}
\begin{align}
x f^{\perp} &=x  \tilde{f}^{\perp}+ f_{1},
\label{eq:EOM1}
&
x h  &=x  \tilde{h} + \frac{k_\perp^2}{M^2}\, h_1^{\perp}\,,
\end{align} 
which allow us to separate the distributions in their twist-2 and
{\em pure} twist-3 ($\tilde{f}^{\perp}$ and $\tilde{h}$)
components. 
Similar equations \cite{Bacchetta:2006tn} hold for the fragmentation functions 
\begin{align}
\frac{\tilde{D}^{\perp}}{z}  &= \frac{D^{\perp}}{z} -  D_1,
\label{e:Dtilde}
&
\frac{\tilde{H}}{z}  &= \frac{H}{z} + \frac{P_{\perp}^2}{M_h^2}\,
H_1^{\perp}.
\end{align}

The perturbative results to first order in $\alpha_s$
for the (subtracted)
functions $f_1$ and $D_1$ 
are well known (see, e.g., 
  Refs.~\cite{Aybat:2011zv,GarciaEchevarria:2011rb,Echevarria:2014rua}): 
\begin{align}
\begin{split}
    {\widehat f }_1^a(x,\bT^2;Q^2,Q^2) & = 
\frac{1}{2\pi}\,\left \{ f_1^a(x,\mu_b^2)  +
\frac{\alpha_s}{\pi}\, \left [  -\frac{1}{4} \, C_F  \left (
  \ln^2\frac{Q^2}{\mu_b^2}  - 3 \ln \frac{Q^2}{\mu_b^2} \right )
  f_1^a(x,\mu_b^2)+  \sum_{i=a,g}\left ( C^{(1)}_{ai} \otimes f_1^i \right
  ) (x,\mu_b^2)   \right ]   \right \} 
\,,
\label{eq:f-lo}
\end{split}
\\
{\widehat D}_1^{a}(z,\bT^2;Q^2,Q^2)  & = \frac{1}{2\pi z^2}\,\left \{
D_1^{a}(z,\mu_b^2)  +  \frac{\alpha_s}{\pi}\, \left [  -\frac{1}{4} \,
  C_F  \left ( \ln^2\frac{Q^2}{\mu_b^2}  - 3 \ln \frac{Q^2}{\mu_b^2}
  \right ) D_1^{a}(z,\mu_b^2) + \sum_{i=a,g}\left ( \hat
  C^{(1)}_{ai}\otimes D_1^{i} \right ) (z,\mu_b^2)   \right ]
\right \} 
\,,
\label{eq:D-lo}
\end{align}
where $\mu_b = 2 e^{-\gamma_E} / \bT$, and $\gamma_E$ is
the Euler constant. 
With this choice, the first-order coefficient functions 
$C^{(1)}$ and $\hat{C}^{(1)}$ become $\bT$-independent.

We then apply the DGLAP equations to
evolve $f_1^a(x,\mu^2) $ from the scale $\mu_b^2$ to
$Q^2$,
\begin{equation}
f_1^a(x,\mu_b^2) = f_1^a(x;Q^2) - \frac{\alpha_s}{2\pi}\, \sum_{i=a,g}(
P_{ai} \otimes f_1^i)(x,Q^2)\,
\ln\frac{Q^2}{\mu_b^2}  + O(\alpha_s^2) \,.
\end{equation}

From this point on, we work out the results only for {\em finite} $k_\T$ and to first order in $\alpha_s$. In this region, and at this order,
we can neglect the contribution coming from the $C^{(1)}$ and $\hat{C}^{(1)}$ coefficient functions.
Using Eqs.~\eqref{eq:int01} and \eqref{eq:int02},
we can Fourier-transform Eqs.~\eqref{eq:f-lo} and \eqref{eq:D-lo} 
and obtain the corresponding results in transverse-momentum space:
\begin{align}
\begin{split}
  f _1^a(x, k_\T^2;Q^2,Q^2)\bigg\vert_{k_\T \ne 0} &=
\frac{\alpha_s}{2\pi^2 k_\T^2}\,\left
\{\frac{1}{2}L\biggl( \frac{Q^2}{k_\T^2}\biggr)  f_1^a(x,Q^2)  +
\sum_{i=a,g}(P_{ai} \otimes f_1^i)(x,Q^2) \right \} \,,
\label{eq:f1-lo-k}
\end{split}
\\
\begin{split}
  D_1^{a}(z, P_\perp^2;Q^2,Q^2)\bigg\vert_{P_\perp \ne 0} & =
\frac{\alpha_s}{2\pi^2 P_\perp^2}\,\left\{
\frac{1}{2} L \biggl( \frac{z^2Q^2}{P_\perp^2}\biggr)  D_1^a(z,Q^2)  +
\sum_{i=a,g}(D_{1}^{i} \otimes P_{ia})(z,Q^2) \right \} \, .
\label{eq:d1-lo-k}
\end{split}
\end{align}

The results of Eqs.~\eqref{eq:f1-lo-k} and \eqref{eq:d1-lo-k} are in agreement with Eqs.~(8.26)
and~(8.47) of Ref.~\cite{Bacchetta:2008xw}, respectively, that were
derived directly in momentum space. 
The only exceptions are the presence of the additional terms $-C_F f_1^a(x)$ 
and $-C_F D_1^a(x)$ in the formulas of Ref.~\cite{Bacchetta:2008xw}, and the different arguments of the logs.
Such a discrepancy is due to the different TMD definitions adopted in the two studies.  
In the present analysis, based on the formalism developed in
Refs.~\cite{Collins:2011zzd,GarciaEchevarria:2011rb,Echevarria:2012js},
the subtracted TMD contains the soft factor that
regulates the rapidity divergences. Conversely, in
Ref.~\cite{Bacchetta:2008xw}  (based on the original CSS
formulation~\cite{Collins:1984kg}) the soft factor is included in
the structure function but not in the unsubtracted TMD itself.   
Including the (square root of the) soft factor in the definition of TMDs removes the rapidity divergences, which in practice reduces to the following replacements w.r.t. the formulas in Ref.~\cite{Bacchetta:2008xw}:
\begin{align}
\label{eq:replacement}
\frac{1}{2}L(\eta^{-1}) &\longrightarrow \frac{1}{2} L\biggl( \frac{Q^2}{k_\T^2}\biggr) + C_F
\,,\nonumber\\
\frac{1}{2}L(\eta_h^{-1}) &\longrightarrow \frac{1}{2} L\biggl( \frac{z^2Q^2}{P_\perp^2}\biggr) + C_F
\,,
\end{align}
where $\eta = k_\perp^2 / x^2 \zeta$, and $\zeta$ is a parameter that defines the gauge-fixing vector in the computation of the quark-quark correlator at twist 3 (see Eqs.~(8.9) and (8.11) in Ref.~\cite{Bacchetta:2008xw}) and regulates the rapidity divergences.
Similarly, $\eta_h = P_\perp^2 / \zeta_h$ (see Eqs.~(8.41) and (8.43) in Ref.~\cite{Bacchetta:2008xw}).
We have verified that, as expected,  the two formalisms lead to the same
unpolarized structure function $F_{UU,T}$ in the intermediate
$q_T$-region.

At high transverse momentum, the chiral-odd functions $h_1^{\perp}$
and $H_1^{\perp}$ are suppressed by factors of $M^2/k_\T^2$ and
$M^2/P_\perp^2$, respectively (see Eq.~(5.45) of
Ref.~\cite{Bacchetta:2008xw}, Eq.~(21) of Ref.~\cite{Zhou:2008fb}, and
Eq.~(6) of Ref.~\cite{Yuan:2009dw}). We will therefore neglect the contributions to Eq.~(\ref{e:F_UUcosphi}) that involve them.

We turn now to the twist-3 chiral-even TMDs, $f^\perp$ and $\tilde
{D}^\perp $. 
The perturbative expansions of their unsubtracted analogues have been calculated at
leading order in momentum space~\cite{Bacchetta:2008xw,Chen:2016hgw}. 
If we apply to these results the same recipe of Eq.~\eqref{eq:replacement} as for the unpolarized TMDs, then from Eq.~(8.27) of
Ref.~\cite{Bacchetta:2008xw} (and also Eq.~(16) of Ref.~\cite{Chen:2016hgw}) we obtain the \emph{subtracted} $f^\perp$:
\begin{align}
x f ^{\perp a}(x, k_\T^2;Q^2,Q^2) \bigg\vert_{k_\T \ne 0}& = \frac{\alpha_s}{4 \pi^2
  k_\T^2}\,\left \{
  {\frac{1}{2} L\biggl( \frac{Q^2}{k_\T^2}\biggr) f_1^a(x,Q^2)  
  + C_F f_1^a(x,Q^2)}
  + \sum_{i=a,g}(P_{ai}' \otimes f_1^i)(x,Q^2) \right \} 
  \,.
\end{align}
Similarly, from Eq.~(8.48) of Ref.~\cite{Bacchetta:2008xw}, the \emph{subtracted} $D^\perp$ is
\begin{align}
\frac{1}{z}\,{D^{ \perp a}}(z, P_\perp^2;Q^2,Q^2) \bigg\vert_{P_\perp \ne 0} =
\frac{\alpha_s}{4\pi^2 P_\perp^2}\, \left \{
   {\frac{1}{2} L\biggl( \frac{z^2Q^2}{P_\perp^2}\biggr) D_1^a(z,Q^2)  
  + C_F D_1^a(z,Q^2)}
+ \sum_{i=a,g} (D_1^i \otimes (2P_{ia}-P^\prime_{ia}))(z,Q^2) \right \}
\,.
\end{align}

Using the EOM relation in Eq.~\eqref{e:Dtilde} we obtain
\begin{align}
  \frac{1}{z}\,{\tilde{D}^{ \perp a}}(z, P_\perp^2;Q^2,Q^2) \bigg\vert_{P_\perp \ne 0} &=
-\frac{\alpha_s}{4\pi^2 P_\perp^2}\, \left\{
   \frac{1}{2} L\biggl( \frac{z^2Q^2}{P_\perp^2}\biggr) D_1^a(z,Q^2)  
  - C_F D_1^a(z,Q^2)
+ \sum_{i=a,g} (D_1^i \otimes  P^\prime_{ia})(z,Q^2) \right \} \, .
\end{align}
Once again, this result differs by the term $- C_F D_1^a$ from Eq.~(8.49) of Ref.~\cite{Bacchetta:2008xw} 
due to the use of
subtracted versus unsubtracted TMDs. 

In order to expand our starting formula in Eq.~(\ref{e:F_UUcosphi}), 
we need the first derivatives of $f^\perp$ and $\tilde D^\perp$ defined 
in Eq.~(\ref{eq:f-der}). Note that any contribution at vanishing
transverse momentum in the original expressions in transverse-momentum space
is irrelevant for these derivatives. Using the integrals in Eqs.~\eqref{eq:int1} 
and \eqref{eq:int2} we find, up to first order in $\alpha_s$, 
\begin{align}
  x {\widehat f }^{\perp (1)\,a}(x,\bT^2;Q^2,Q^2)
  & =  \frac{1}{M^2\bT^2} \, \frac{\alpha_s}{4\pi^2}\,\left [
   {\frac{1}{2} L\biggl( \frac{Q^2}{\mu_b^2}\biggr) f_1^a(x,Q^2)  
  + C_F f_1^a(x,Q^2)}
    +\sum_i\left (P^\prime_{a i}\otimes f_1^i \right )
    (x,Q^2) \right ] 
\,,
\label{eq:f-der1} \\
 \frac{1}{z}\,{\widehat {\tilde D} }^{ \perp (1)\,a }(z,\bT^2;Q^2,Q^2)
  &= - \frac{1}{z^2M_h^2\bT^2} \, \frac{\alpha_s}{4\pi^2} \left [
    {\frac{1}{2} L\biggl( \frac{Q^2}{\mu_b^2}\biggr) D_1^a(z,Q^2)  
  - C_F D_1^a(z,Q^2)}
   +  \sum_i \left (D_1^{i} \otimes
   P^\prime_{ai} \right ) (z, Q^2) \right ] 
\,.
\label{eq:D-der1}
\end{align}

If in Eq.~(\ref{e:F_UUcosphi}) we are only interested in the high-$q_{T}$ behavior of the structure functions and we work at first order in $\alpha_s$, the $\mathcal{H}'_{\mathrm{SIDIS}}$ factor can be set equal to 1. By further inserting the expressions of $\widehat{f}_1^a$ and $\widehat{D}_1^{a}$ from Eqs.~(\ref{eq:f-lo}, \ref{eq:D-lo}), respectively, the expressions of $\widehat{f}^{\perp (1) a}$ and $\smash{\widehat{\tilde{D}}}^{\perp  (1) a}$ from Eqs.~(\ref{eq:f-der1}, \ref{eq:D-der1}), respectively, and further neglecting the suppressed contributions from chiral-odd TMDs, we get
\begin{align}
\begin{split}
F_{UU}^{\cos\phi_h}  & = - \frac{4 \pi M^2}{Q }\, {\Bigg\{ } \sum_a e_a^2 x
\int_0^\infty d\bT\,\bT^2\, J_1 \biggl( \frac{\bT P_{\T}}{z} \biggr)
\left ( - \frac{M_h^2}{M^2}\,\widehat{f}_1^a
\frac{\smash{\widehat{\tilde{D}}}^{\perp (1) a}}{z}  +  x
\widehat{f}^{\perp (1) a} \widehat{D}_1^{a}  \right )
+{\cal O} \biggl( \frac{\Lambda_{\rm QCD}}{q_T} \biggr) + {\cal O} \biggl( \frac{q_T}{Q} \biggr) {\Bigg\} }
\\ & = - \frac{1}{Q}\, \frac{\alpha_s}{2\pi^2 z^2} {\Bigg\{ } \sum_a  e_a^2 x
\int_0^\infty d\bT \, J_1 \biggl ( \frac{\bT P_{\T}}{z} \biggr ) \left
    [  L\biggl(\frac{Q^2}{\mu_b^2} \biggr)
      f_1^a(x,Q^2)\,D_1^{a}(z, Q^2) \right .  \\ & \quad +
      \sum_{i=a,g} \biggl(f_1^a(x; Q^2) (D_1^i \otimes
  P_{ia}')(z; Q^2) + (P_{ai}' \otimes f_{1}^{i})(x; Q^2)\, D_1^a(z; Q^2)\biggr)\biggr] 
  +{\cal O} \biggl( \frac{\Lambda_{\rm QCD}}{q_T} \biggr) + {\cal O} \biggl( \frac{q_T}{Q} \biggr) {\Bigg\} }
  \,.
\label{e:lowFUUcosphi}
\end{split}
\end{align}

Finally, using again the integrals of Eqs.~\eqref{eq:int1} and \eqref{eq:int2}
we recover the dominant term of $F_{UU}^{\cos\phi_h}$ in the intermediate momentum region,
\begin{align}
\begin{split}
\label{e:low_FUUcosphi}
F_{UU}^{\cos\phi_h} &= - \frac{1}{Q q_T}\, \frac{\alpha_s}{2\pi^2 z^2} {\Bigg\{ }
\sum_a x e_a^2\, \biggl[ L \biggl(
  \frac{Q^2}{q_T^2} \biggr) f_1^a(x, Q^2) \, D_1^a(z, Q^2)
     \\ & \quad
  +       \sum_{i=a,g} \biggl(f_1^a(x; Q^2) (D_1^i \otimes
  P_{ia}')(z; Q^2) + 
 (P_{ai}' \otimes f_{1}^{i})(x; Q^2)\, D_1^a(z; Q^2)\biggr)\biggr]
     \\ & \quad
   +{\cal O}\biggl (
\frac{\Lambda_{\rm QCD}}{q_T}\biggr )+{\cal O}\biggl (
\frac{q_T}{Q}\biggr ) {\Bigg\} } \,.
\end{split}
\end{align}
This result is identical to the one in Eq.~(\ref{e:high_FUUcosphi})
obtained in the collinear framework. On the contrary, in Ref.~\cite{Bacchetta:2008xw} 
a mismatch was found between the two descriptions at high- and low-$q_T$ 
because of the extra term $-2C_F f_1^a  D_1^a$ appearing only in the latter. 
We deduce that a systematic matching between the two descriptions (not only for  
$F_{UU,T}$ but also for $F_{UU}^{\cos\phi_h}$) 
is possible only by adopting the TMD definition of 
Refs.~\cite{Collins:2011zzd,GarciaEchevarria:2011rb,Echevarria:2012js}, 
that directly includes the (square root of the) soft factor. This applies to all the TMDs, not only to the unpolarized ones.
\section{$\cos\phi$ asymmetry in Drell-Yan}
\label{s:dy}

Along the lines of the previous section, we now study the unpolarized DY process,
\begin{equation}
h_1 (P_1) + h_2(P_2) \to \ell(l) + \bar{\ell}(l^\prime) + X\,,
\end{equation}
where the momenta of the particles are within brackets. We consider
only the electromagnetic interaction and denote by $q=l+l^\prime$ the
4-momentum of the exchanged virtual photon with $Q^2=q^2$, being $q_T$
its transverse component orthogonal to $P_1$ and $P_2$. The
angular dependence of the cross section is conveniently written in the
dilepton rest frame~\cite{Arnold:2008kf,Lu:2011mz,Lu:2011th},
\begin{align}
\frac{d\sigma}{d^4 q \,d\Omega} = \frac{\alpha^2}{2 s Q^2}\, \left \{
(1+\cos^2\theta)  F^1_{UU} + (1-\cos^2\theta)  F^2_{UU} + \sin 2\theta
\cos\phi\, F_{UU}^{\cos\phi} + \sin^2\theta\cos 2\phi   F_{UU}^{\cos 2
  \phi}\right \} \,,
\end{align} 
where $d\Omega = d\cos\theta d\phi$ is the solid angle of the lepton
$\ell$ and $s = (P_1+P_2)^2$.
Measurements related to the above structure functions
have been presented in
in Refs.~\cite{Guanziroli:1987rp,Conway:1989fs,Zhu:2006gx,Zhu:2008sj,
  Aaltonen:2011nr,Khachatryan:2015paa,Khachatryan:2015paa,Aad:2016izn}. 
Phenomenological analyses have been reported in, e.g.,
Refs.~\cite{Barone:2010gk,Peng:2015spa,Lambertsen:2016wgj,Motyka:2016lta,Peng:2018tty}.

In the literature there are two common 
choices of reference frames, depending on the choice of the $\hat z$ axis: in the 
Collins-Soper frame (CS)~\cite{Collins:1977iv}, the $\hat z$ axis 
points in the direction that bisects the angle between $\bm{P}_1$
and $-\bm{P}_2$; in the Gottfried-Jackson frame
(GJ)~\cite{Lam:1978pu}, it points in the direction of $\bm{P}_1$. A
linear transformation connects the structure functions in the two frames~\cite{Boer:2006eq},
\begin{equation} 
\left( \begin{array}{c} F_{UU}^1 \\[3pt] F_{UU}^2 \\[3pt]
  F_{UU}^{\cos\phi} \\[3pt] F_{UU}^{\cos2\phi}
  \end{array} \right)_{\rm GJ}
= \frac{1}{1+\rho^2} \left( \begin{array}{cccc} 1+\frac{1}{2}\rho^2 &
  \frac{1}{2} \rho^2 & - \rho & \frac{1}{2}\rho^2 \\[3pt] \rho^2 & 1 &
  2\rho & -\rho^2 \\[3pt] \rho & - \rho & 1- \rho^2 & -\rho \\[3pt]
  \frac{1}{2} \rho^2 & - \frac{1}{2} \rho^2 & \rho & 1+
  \frac{1}{2}\rho^2 
\end{array} \right) \; 
\left( \begin{array}{c}  F_{UU}^1 \\[3pt] F_{UU}^2 \\[3pt]
  F_{UU}^{\cos\phi} \\[3pt] F_{UU}^{\cos2\phi}
\end{array} \right)_{\rm CS} \, ,
\label{fromCStoGJ}
\end{equation} 
where $\rho = q_T/Q$.

In collinear factorization, the structure functions $F_{UU}^1$ and
$F_{UU}^{\cos\phi}$
can be calculated to ${\cal O} (\alpha_s)$ in the intermediate
transverse-momentum region $(\Lambda_{\rm QCD }\ll q_T \ll Q)$ in both
frames (see Eqs.~(33,37) of Ref.~\cite{Boer:2006eq}):\footnote{The prefactors
  in our formulas are different from Ref.~\cite{Boer:2006eq} due to different
  definitions of the structure functions.}
\begin{align}
\begin{split}
\label{eq:FUU1high}
{F_{UU}^1} \Big\vert_{\rm GJ} =  {F_{UU}^1} \Big\vert_{\rm CS}   &=  \frac{\alpha_s}{2 \pi^2
  q_T^2} {\Bigg\{ } \sum_a \frac{e_a^2}{N_c} \biggl[
  L\biggl(\frac{Q^2}{q_T^2} \biggr)
  f_1^a(x_1) f_1^{\bar a}(x_2)
  \\ & \quad + \sum_{i=a,g}\biggl((P_{ai} \otimes
  f_1^{i}) (x_1) f_1^{\bar a}(x_2) 
  + f_1^a (x_1)(P_{ai} \otimes f_1^{i}) (x_2)\biggl) \biggr]
+{\cal O}\biggl (
\frac{\Lambda_{\rm QCD}}{q_T}\biggr )+{\cal O}\biggl (
\frac{q_T}{Q}\biggr ) {\Bigg\} }
\,,
\end{split}
\\
\begin{split}
\label{eq:F-GJ}
{F_{UU}^{\cos\phi}} \Big\vert_{\rm GJ}   &=  \frac{\alpha_s}{2 \pi^2 Q
  q_T} {\Bigg\{ } \sum_a \frac{e_a^2}{N_c} \biggl[
  L\biggl(\frac{Q^2}{q_T^2} \biggr)
  f_1^a(x_1) f_1^{\bar a}(x_2)
\\ & \quad  + \sum_{i=a,g}\biggl((P_{ai}^\prime \otimes
  f_1^{i}) (x_1) f_1^{\bar a}(x_2) 
  + f_1^a (x_1)((2P_{ai}-P_{ai}^\prime) \otimes f_1^{i}) (x_2)\biggl) \biggr]
  +{\cal O}\biggl (
  \frac{\Lambda_{\rm QCD}}{q_T}\biggr )
  +{\cal O}\biggl(\frac{q_T}{Q}\biggr) {\Bigg\} } 
  \, ,
\end{split}
\\
\begin{split}
  \label{eq:F-CS}
{F_{UU}^{\cos\phi}} \Big\vert_{\rm CS}  &= - \frac{\alpha_s}{2 \pi^2 Q q_T} {\Bigg\{ } 
\sum_a \frac{e_a^2}{N_c} 
  \sum_{i=a,g}\biggl[((P_{ai}-P_{ai}^\prime)
  \otimes f_1^{i})(x_1) f_1^{\bar a}(x_2)
  -  f_1^a (x_1) ((P_{ai}-P_{ai}^\prime) \otimes f_1^{i}) (x_2)
  \biggr]
  \\ & \quad
  +{\cal O}\biggl(
\frac{\Lambda_{\rm QCD}}{q_T}\biggr)+{\cal O}\biggl(
\frac{q_T}{Q}\biggr) {\Bigg\} } 
  \,,
\end{split}
\end{align}
where for simplicity we have suppressed the dependence of PDFs on the hard scale $Q^{2}$.
Higher-order contributions in $\alpha_s$ have been implemented in
Ref.~\cite{Lambertsen:2016wgj,Gauld:2017tww}. Strikingly,  $F_{UU}^{\cos\phi}$ have very different behaviors in CS and CJ frames, namely, the logarithmic terms are not present in CS frame. This fact was the reason to believe that resummation of $F_{UU}^{\cos\phi}$ is very different from CSS one (see Ref.~\cite{Boer:2006eq}). The authors of Ref.~\cite{Berger:2007jw} pointed out that the formalism based on collinear QCD factorization is not enough  to reconcile these behaviors and derive the correct resummed formula. In order to obtain the complete resummation result, we will start from the TMD expression and show how it reduces to the correct collinear expressions in both frames.

In the low-$q_T$ region, coherently with Eq.~\eqref{e:F_UUcosphi},
we {\em assume}
that the parton-model result for the $F_{UU}^{\cos\phi}$ structure function (see, e.g., Ref.~\cite{Lu:2011th})  
can be generalized by including higher-order contributions to the hard scattering and
replacing the TMDs with the subtracted ones. We obtain then 
\begin{align} 
\begin{split}
F_{UU}^{1} & = \mathcal{H}_{\rm DY}  \, \sum_a \frac{e_a^2}{N_c}\,
  \mathcal{B}_0 \Bigl[ \widehat{f}_1^{a}
    \widehat{f}_1^{\bar{a}}\Bigr]
  +{\cal O}\left(\frac{q_{T}^2}{Q^2}\right)\, ,
\end{split} 
\\
\begin{split}
  F_{UU}^{\cos\phi} & = \frac{2 M^2}{Q}  {\Bigg\{ }  \mathcal{H}'_{\rm DY} \, \sum_a \frac{e_a^2}{N_c}\,
  \mathcal{B}_1 \Biggl[ \biggl( (1-c) x_1
    \widehat{f}^{\perp (1)\, a} + c x_1 \widehat{\tilde{f}}^{\perp
      (1)\, a}  \biggr) \widehat{f}_1^{\, \bar{a}} -
  \widehat{f}_1^{\,a} \biggl( c \,x_2
    \widehat{f}^{\,\perp\,(1)\,\bar{a}} + (1-c)\,x_2 \widehat{\tilde
      f}^{\,\perp\,(1)\,\bar{a}} \biggr)
  \\
  & \quad +\frac{M_2}{M_1}
    \widehat{h}_1^{\perp (1) a} \biggl(c\,x_2 \widehat{h}^{\bar{a}}+(1-c)\,x_2
    \widehat{\tilde{h}}^{\bar{a}}\biggr)
  - \frac{M_1}{M_2}\biggl(
    \Big((1-c)\,x_1 \widehat{h}^{a}+c\,x_1
    \widehat{\tilde{h}}^{a}\Big) \widehat{h}_1^{\perp (1) \bar{a}} \biggr)
    \Biggr]
  +{\cal O}\biggl(\frac{q_{T}}{Q}\biggr) {\Bigg\} }
  \, ,
  \end{split}
\label{eq:FcosphiDY}
\end{align} 
where
$c=0$ for the GJ frame, $c=1/2$ for the CS frame, and $M_i$, with
$i=1,2$,  are the masses of the initial hadrons.

It is important to note that since our formula contains subtracted TMDs,
we can guarantee the validity of the relations of Eq.~\eqref{fromCStoGJ} connecting different frames.
In fact, the difference between the GJ and CS frames is, at first order in $\rho$, 
\begin{equation}
  {F_{UU}^{\cos\phi}} \Big\vert_{\rm GJ} - {F_{UU}^{\cos\phi}} \Big\vert_{\rm CS}
  = \rho \Big( {F_{UU}^1} - F_{UU}^{\cos 2 \phi} \Big)\Big\vert_{\rm CS}
  + {\cal O}\big(\rho^{{2}} \big).
\end{equation}
The structure functions on the right-hand side (at this order, it is not relevant whether they are written in one frame or the other) 
contribute to the leading-twist part of the cross section. For the factorization theorem to hold, they must contain subtracted TMDs. We have directly checked that the difference of subleading-twist structure functions on the left-hand side matches the right-hand side only if subtracted TMDs are involved, as we conjecture in Eq.~(\ref{eq:FcosphiDY}).

The dominant contribution of $F_{UU}^{\cos\phi}$ in the intermediate
$q_T$ region can now be calculated in a straightforward way, along the
lines of the previous section. We neglect the  (suppressed) chiral-odd terms.
We replace $\widehat{f}_1$ with the expression in Eq.~(\ref{eq:f-lo}), 
$\widehat{f}^{\,\perp \,(1)}$ with Eq.~(\ref{eq:f-der1}), and we obtain
$\widehat{\tilde  f}^{\,\perp \,(1)}$ from the relation 
$x \widehat{\tilde f}^{\,\perp \,(1)} =  x\widehat{f}^{\,\perp\,(1)} 
- \widehat{f}_1^{(1)}$, as in Eq.~(\ref{eq:EOM1}). We also use the following formula
\begin{align}
  \begin{split}
{\widehat f }_1^{(1)\,a}(x,\bT^2;Q^2,Q^2) & \equiv  -\frac{1}{M^2}\,\frac{1}{\bT}\,\frac{\partial}{\partial \bT}\,
{\widehat f }_1^{a}(x,\bT^2;Q^2,Q^2)
\\ &
= \frac{1}{M^2\bT^2} \, \frac{\alpha_s}{4\pi^2}\biggl[ L\biggl(\frac{Q^2}{\mu_b^2} \biggr) f_1^a(x,Q^2) + 
2 \sum_{i=a,g} (P_{ai} \otimes f_1^i) (x,Q^2) \biggr] \, .
\label{eq:f1-der1}
 \end{split}
\end{align}

By performing these substitutions, we obtain the general expression 
\begin{align}
\begin{split}
F_{UU}^{\cos\phi}
& = \frac{\alpha_s}{2 \pi^2 Q q_T} {\Bigg\{ } \sum_a \frac{e_a^2}{N_c} 
\Biggl\{ (1-2c) \,  L\biggl(\frac{Q^2}{q_T^2} \biggr)
  f_1^a(x_1,Q^2) f_1^{\bar a}(x_2,Q^2)
\\ & \quad
  + \sum_{i=a,g}\biggl[ \bigl[ (-2cP_{ai}+ P_{ai}^\prime) \otimes
  f_1^{i} \bigr] (x_1,Q^2) f_1^{\bar a}(x_2,Q^2)
\\ & \quad 
+ f_1^a (x_1,Q^2) \bigl[ \bigl( (2-2c)P_{ai}-P_{ai}^\prime \bigr) \otimes f_1^{i} \bigr] (x_2,Q^2)
\biggr] \Biggr\} 
+{\cal O}\biggl (
\frac{\Lambda_{\rm QCD}}{q_T}\biggr )
+{\cal O}\biggl(\frac{q_T}{Q}\biggr) {\Bigg\} } 
\, .
\label{eq:FcosphiDY-2}
\end{split}
\end{align}
It can be verified that the above expression is in agreement with
Eq.~(\ref{eq:F-GJ}) for $c=0$ and with Eq.~\eqref{eq:F-CS} for $c=1/2$. We have reproduced the correct large-$q_T$ results 
in both frames that were obtained in Refs.~\cite{Boer:2006eq,Berger:2007jw}, and thus have solved the long standing problem of resummation of collinear results in large $q_T$ region.

In the so-called Wandzura--Wilczek approximation, the pure twist-three
components (functions with tilde) are neglected. In
App.~\ref{a:WW} we show that also in this case the leading logarithmic term proportional to $L(Q^2/q_T^2)$ matches
in the two descriptions at low and high $q_T$, when using subtracted TMDs.
The nonlogarithmic terms however cannot be correctly reproduced in Wandzura--Wilczek approximation.

\section{Conclusions}

In this work, we analyzed the $\cos \phi$ modulation of the unpolarized
cross section in semi-inclusive DIS (SIDIS) and in the Drell-Yan process (DY). 
At high values of transverse momentum in the final-state system, this
observable can be computed in a standard way in terms of collinear
unpolarized PDF and FF. 
At low transverse momentum, it represents the simplest
observable that can be written in terms of subleading-twist TMDs.
However, no factorization proof for TMD
observables at twist-3 is available 
and inconsistencies have been pointed out in the literature~\cite{Boer:2006eq,Berger:2007jw,Bacchetta:2008xw,Chen:2016hgw},
casting a doubt on the possibility of achieving such a proof.

For the low transverse momentum region, in analogy with twist-2 observables,
we propose a simple modification of the parton-model
formula, with the replacement of TMDs with subtracted TMDs, according to 
Refs.~\cite{Collins:2011zzd,GarciaEchevarria:2011rb,Echevarria:2012js}.
We show  
that our formula correctly matches the
collinear result at high transverse momentum for both SIDIS and DY, solving a problem
first highligthed in Ref.~\cite{Bacchetta:2008xw}.

As for DY, we further show that
our formula guarantees the correct behavior
under the change of frames of reference, which
is a nontrivial feature when considering twist-3 contributions (see also
Ref.~\cite{Kanazawa:2015ajw}). We also solve the long standing problem of resummation of collinear QCD results in large--$q_T$ region posed in Refs.~\cite{Boer:2006eq,Berger:2007jw}.

Our conjecture provides a formula for the $\cos \phi$ modulation at low
transverse momentum that is compatible with TMD factorization  
up to subleading twist.  It can be readily applied to other modulations and to electron-positron annihilation~\cite{Boer:2008fr}. 
We believe that this is an important step towards the full proof. 

\begin{acknowledgments}
We thank D.~Boer and M.~Diehl for useful discussions. This work is
supported by the European Research Council (ERC) under the European
Union's Horizon 2020 research and innovation program (grant agreement
No. 647981, 3DSPIN). 
MGE is supported by the Marie Sk\l odowska-Curie grant \emph{GlueCore} (grant agreement No. 793896). This work is supported by the U.S.\
Department of Energy under Contract No.~DE-AC05-06OR23177 
and within the TMD Collaboration framework, and by the National
Science Foundation under Contract No.\ PHY-1623454.
\end{acknowledgments}

\appendix

\section{Bessel integrals}
\label{a:besselint}

Using the notation $b_{0}=2\exp(-\gamma_{E})$, we can write the following
integrals 

\begin{align}
\label{eq:int01}
&\int_0 ^\infty d x \, x \, J_0 \left (x y  \right ) \ln
\frac{A^2 x^{2}}{b_{0}^2} \ \Biggl|_{y\neq 0}
= -\frac{2}{y^2},\\ 
\label{eq:int02}
&\int_0 ^\infty d x \, x \, J_0 \left (x y  \right ) \ln^2
\frac{A^2x^{2}}{b_{0}^2}\ \Biggl|_{y\neq 0}
= -\frac{4}{y^2}  \,\ln \frac{A^2}{y^2},\\
\label{eq:int1}
&\int_0 ^\infty d x \, J_1 \left (x y  \right ) \ln
\frac{A^2x^{2}}{b_{0}^2} = \frac{1}{y} \,\ln \frac{A^2}{y^2},\\
\label{eq:int2}
&\int_0 ^\infty d x \, J_n \left (x y   \right ) =
\frac{1}{y}.
\end{align}

\section{Wandzura--Wilczek approximation}
\label{a:WW}

In the so-called Wandzura--Wilczek approximation, the pure twist-three
components of TMDs (functions with tilde) are neglected. The EOM relations in Eqs.~(\ref{eq:EOM1}--\ref{e:Dtilde}) reduce to  
\begin{equation}
xf^\perp \approx f_1 \, , \qquad xh \approx \frac{k_\perp^2}{M^2} \,
h_1^\perp \, , \qquad D^\perp \approx z D_1 \, , \qquad H \approx -z
\frac{P_{\perp}^2}{M_h^2} \, H_1^{\perp} \, .   
\end{equation}

The formula for the structure function $F_{UU}^{\cos\phi_h}$ in SIDIS,
Eq.~\eqref{e:F_UUcosphi}, considerably simplifies and we obtain (neglecting chiral-odd terms)
\begin{align}
\begin{split}
  F_{UU}^{\cos\phi_h}  & \stackrel{\text{\tiny WW}}{=}
    - \frac{2 M^2}{Q}  \mathcal{H}'_{\rm {SIDIS}} \, \sum_a \, e_a^2 x  \mathcal{B}_1 \Bigl[
\widehat{f}_1^{(1) a} \widehat{D}_1^{a} \Bigr]
\\ & \approx - \frac{1}{Q q_T}\, \frac{\alpha_s}{2\pi^2 z^2} {\Bigg\{ } 
\sum_a x e_a^2\, \biggl[L \biggl(
  \frac{Q^2}{q_T^2} \biggr) f_1^a(x, Q^2) \, D_1^a(z, Q^2)
\\ & \quad
  + 2\sum_{i=a,g}( P_{ai} \otimes f_1^i) (x, Q^2) \, D_1^a(z, Q^2)
  \biggr]
+{\cal O}\biggl(\frac{\Lambda_{\rm QCD}}{P_{hT}}\biggr)+{\cal O}\biggl(\frac{P_{hT}}{Q}\biggr) {\Bigg\} . }
\end{split}
\end{align}
Comparing this result with Eq.~\eqref{e:high_FUUcosphi}, we see that the 
leading logarithmic term proportional to $L(Q^2/q_T^2)$ matches
in the two descriptions at low and high $q_T$, when using subtracted TMDs.
The nonlogarithmic terms are not correctly reproduced
in the Wandzura--Wilczek approximation.

Similarly, for Drell--Yan we obtain
\begin{equation}
  \begin{split}
    F_{UU}^{\cos\phi} & \stackrel{\text{\tiny WW}}{=}
   \frac{2 M^2}{Q} \mathcal{H}'_{\rm DY} \, \sum_a \frac{e_a^2}{N_c}\,
   \mathcal{B}_1  \biggl[ (1-c)
    \widehat{f}_1^{(1)\, a} \widehat{f}_1^{\, \bar{a}} - c
  \widehat{f}_1^{\,a} 
  \widehat{f}_1^{(1)\,\bar{a}}  \biggr]
  \\
  & \approx \frac{\alpha_s}{2 \pi^2 Q q_T} {\Biggl\{ } \sum_a \frac{e_a^2}{N_c}
  \Biggl[ (1-2c)
  L\biggl(\frac{Q^2}{q_T^2} \biggr)
  f_1^a(x_1,Q^2) f_1^{\bar a}(x_2,Q^2)
\\ & \quad
  + \sum_{i=a,g}\biggl[2(1-c)(P_{ai} \otimes
  f_1^{i}) (x_1,Q^2) f_1^{\bar a}(x_2,Q^2)
\\ & \quad 
-2c  f_1^a (x_1,Q^2)(P_{ai} \otimes f_1^{i}) (x_2,Q^2)\biggr]
  \Biggr]
  +{\cal O}\biggl(\frac{\Lambda_{\rm QCD}}{q_{T}}\biggr)+{\cal  O}\biggl(\frac{q_{T}}{Q}\biggr) {\Biggr\} }
  \, ,
  \end{split}
\end{equation}
where only the leading logarithmic term proportional to $L (Q^2/q_T^2)$ matches to Eq.~(\ref{eq:FcosphiDY-2}).

\bibliographystyle{apsrevM}
\bibliography{cosphibiblio}

\ifx\mcitethebibliography\mciteundefinedmacro
\PackageError{apsrevM.bst}{mciteplus.sty has not been loaded}
{This bibstyle requires the use of the mciteplus package.}\fi
\begin{mcitethebibliography}{60}
\expandafter\ifx\csname natexlab\endcsname\relax\def\natexlab#1{#1}\fi
\expandafter\ifx\csname bibnamefont\endcsname\relax
  \def\bibnamefont#1{#1}\fi
\expandafter\ifx\csname bibfnamefont\endcsname\relax
  \def\bibfnamefont#1{#1}\fi
\expandafter\ifx\csname citenamefont\endcsname\relax
  \def\citenamefont#1{#1}\fi
\expandafter\ifx\csname url\endcsname\relax
  \def\url#1{\texttt{#1}}\fi
\expandafter\ifx\csname urlprefix\endcsname\relax\def\urlprefix{URL }\fi
\providecommand{\bibinfo}[2]{#2}
\providecommand{\eprint}[2][]{\url{#2}}

\bibitem[{\citenamefont{Collins}(2011)}]{Collins:2011zzd}
\bibinfo{author}{\bibfnamefont{J.}~\bibnamefont{Collins}},
  \bibinfo{journal}{Camb. Monogr. Part. Phys. Nucl. Phys. Cosmol.}
  \textbf{\bibinfo{volume}{32}}, \bibinfo{pages}{1}
  (\bibinfo{year}{2011})\relax
\mciteBstWouldAddEndPuncttrue
\mciteSetBstMidEndSepPunct{\mcitedefaultmidpunct}
{\mcitedefaultendpunct}{\mcitedefaultseppunct}\relax
\EndOfBibitem
\bibitem[{\citenamefont{Echevarria et~al.}(2012)\citenamefont{Echevarria,
  Idilbi, and Scimemi}}]{GarciaEchevarria:2011rb}
\bibinfo{author}{\bibfnamefont{M.~G.} \bibnamefont{Echevarria}},
  \bibinfo{author}{\bibfnamefont{A.}~\bibnamefont{Idilbi}}, \bibnamefont{and}
  \bibinfo{author}{\bibfnamefont{I.}~\bibnamefont{Scimemi}},
  \bibinfo{journal}{JHEP} \textbf{\bibinfo{volume}{07}}, \bibinfo{pages}{002}
  (\bibinfo{year}{2012}), \eprint{1111.4996}\relax
\mciteBstWouldAddEndPuncttrue
\mciteSetBstMidEndSepPunct{\mcitedefaultmidpunct}
{\mcitedefaultendpunct}{\mcitedefaultseppunct}\relax
\EndOfBibitem
\bibitem[{\citenamefont{Echevarría et~al.}(2013)\citenamefont{Echevarría,
  Idilbi, and Scimemi}}]{Echevarria:2012js}
\bibinfo{author}{\bibfnamefont{M.~G.} \bibnamefont{Echevarría}},
  \bibinfo{author}{\bibfnamefont{A.}~\bibnamefont{Idilbi}}, \bibnamefont{and}
  \bibinfo{author}{\bibfnamefont{I.}~\bibnamefont{Scimemi}},
  \bibinfo{journal}{Phys. Lett.} \textbf{\bibinfo{volume}{B726}},
  \bibinfo{pages}{795} (\bibinfo{year}{2013}), \eprint{1211.1947}\relax
\mciteBstWouldAddEndPuncttrue
\mciteSetBstMidEndSepPunct{\mcitedefaultmidpunct}
{\mcitedefaultendpunct}{\mcitedefaultseppunct}\relax
\EndOfBibitem
\bibitem[{\citenamefont{Bacchetta et~al.}(2008)\citenamefont{Bacchetta, Boer,
  Diehl, and Mulders}}]{Bacchetta:2008xw}
\bibinfo{author}{\bibfnamefont{A.}~\bibnamefont{Bacchetta}},
  \bibinfo{author}{\bibfnamefont{D.}~\bibnamefont{Boer}},
  \bibinfo{author}{\bibfnamefont{M.}~\bibnamefont{Diehl}}, \bibnamefont{and}
  \bibinfo{author}{\bibfnamefont{P.~J.} \bibnamefont{Mulders}},
  \bibinfo{journal}{JHEP} \textbf{\bibinfo{volume}{08}}, \bibinfo{pages}{023}
  (\bibinfo{year}{2008}), \eprint{0803.0227}\relax
\mciteBstWouldAddEndPuncttrue
\mciteSetBstMidEndSepPunct{\mcitedefaultmidpunct}
{\mcitedefaultendpunct}{\mcitedefaultseppunct}\relax
\EndOfBibitem
\bibitem[{\citenamefont{Collins et~al.}(1985)\citenamefont{Collins, Soper, and
  Sterman}}]{Collins:1984kg}
\bibinfo{author}{\bibfnamefont{J.~C.} \bibnamefont{Collins}},
  \bibinfo{author}{\bibfnamefont{D.~E.} \bibnamefont{Soper}}, \bibnamefont{and}
  \bibinfo{author}{\bibfnamefont{G.~F.} \bibnamefont{Sterman}},
  \bibinfo{journal}{Nucl. Phys.} \textbf{\bibinfo{volume}{B250}},
  \bibinfo{pages}{199} (\bibinfo{year}{1985})\relax
\mciteBstWouldAddEndPuncttrue
\mciteSetBstMidEndSepPunct{\mcitedefaultmidpunct}
{\mcitedefaultendpunct}{\mcitedefaultseppunct}\relax
\EndOfBibitem
\bibitem[{\citenamefont{Catani et~al.}(2001)\citenamefont{Catani, de~Florian,
  and Grazzini}}]{Catani:2000vq}
\bibinfo{author}{\bibfnamefont{S.}~\bibnamefont{Catani}},
  \bibinfo{author}{\bibfnamefont{D.}~\bibnamefont{de~Florian}},
  \bibnamefont{and} \bibinfo{author}{\bibfnamefont{M.}~\bibnamefont{Grazzini}},
  \bibinfo{journal}{Nucl. Phys.} \textbf{\bibinfo{volume}{B596}},
  \bibinfo{pages}{299} (\bibinfo{year}{2001}), \eprint{hep-ph/0008184}\relax
\mciteBstWouldAddEndPuncttrue
\mciteSetBstMidEndSepPunct{\mcitedefaultmidpunct}
{\mcitedefaultendpunct}{\mcitedefaultseppunct}\relax
\EndOfBibitem
\bibitem[{\citenamefont{Gamberg et~al.}(2018)\citenamefont{Gamberg, Metz,
  Pitonyak, and Prokudin}}]{Gamberg:2017jha}
\bibinfo{author}{\bibfnamefont{L.}~\bibnamefont{Gamberg}},
  \bibinfo{author}{\bibfnamefont{A.}~\bibnamefont{Metz}},
  \bibinfo{author}{\bibfnamefont{D.}~\bibnamefont{Pitonyak}}, \bibnamefont{and}
  \bibinfo{author}{\bibfnamefont{A.}~\bibnamefont{Prokudin}},
  \bibinfo{journal}{Phys. Lett.} \textbf{\bibinfo{volume}{B781}},
  \bibinfo{pages}{443} (\bibinfo{year}{2018}), \eprint{1712.08116}\relax
\mciteBstWouldAddEndPuncttrue
\mciteSetBstMidEndSepPunct{\mcitedefaultmidpunct}
{\mcitedefaultendpunct}{\mcitedefaultseppunct}\relax
\EndOfBibitem
\bibitem[{\citenamefont{Echevarria et~al.}(2018)\citenamefont{Echevarria,
  Kasemets, Lansberg, Pisano, and Signori}}]{Echevarria:2018qyi}
\bibinfo{author}{\bibfnamefont{M.~G.} \bibnamefont{Echevarria}},
  \bibinfo{author}{\bibfnamefont{T.}~\bibnamefont{Kasemets}},
  \bibinfo{author}{\bibfnamefont{J.-P.} \bibnamefont{Lansberg}},
  \bibinfo{author}{\bibfnamefont{C.}~\bibnamefont{Pisano}}, \bibnamefont{and}
  \bibinfo{author}{\bibfnamefont{A.}~\bibnamefont{Signori}},
  \bibinfo{journal}{Phys. Lett.} \textbf{\bibinfo{volume}{B781}},
  \bibinfo{pages}{161} (\bibinfo{year}{2018}), \eprint{1801.01480}\relax
\mciteBstWouldAddEndPuncttrue
\mciteSetBstMidEndSepPunct{\mcitedefaultmidpunct}
{\mcitedefaultendpunct}{\mcitedefaultseppunct}\relax
\EndOfBibitem
\bibitem[{\citenamefont{Boer and Vogelsang}(2006)}]{Boer:2006eq}
\bibinfo{author}{\bibfnamefont{D.}~\bibnamefont{Boer}} \bibnamefont{and}
  \bibinfo{author}{\bibfnamefont{W.}~\bibnamefont{Vogelsang}},
  \bibinfo{journal}{Phys. Rev.} \textbf{\bibinfo{volume}{D74}},
  \bibinfo{pages}{014004} (\bibinfo{year}{2006}), \eprint{hep-ph/0604177}\relax
\mciteBstWouldAddEndPuncttrue
\mciteSetBstMidEndSepPunct{\mcitedefaultmidpunct}
{\mcitedefaultendpunct}{\mcitedefaultseppunct}\relax
\EndOfBibitem
\bibitem[{\citenamefont{Berger et~al.}(2007)\citenamefont{Berger, Qiu, and
  Rodriguez-Pedraza}}]{Berger:2007jw}
\bibinfo{author}{\bibfnamefont{E.~L.} \bibnamefont{Berger}},
  \bibinfo{author}{\bibfnamefont{J.-W.} \bibnamefont{Qiu}}, \bibnamefont{and}
  \bibinfo{author}{\bibfnamefont{R.~A.} \bibnamefont{Rodriguez-Pedraza}},
  \bibinfo{journal}{Phys. Rev.} \textbf{\bibinfo{volume}{D76}},
  \bibinfo{pages}{074006} (\bibinfo{year}{2007}), \eprint{0708.0578}\relax
\mciteBstWouldAddEndPuncttrue
\mciteSetBstMidEndSepPunct{\mcitedefaultmidpunct}
{\mcitedefaultendpunct}{\mcitedefaultseppunct}\relax
\EndOfBibitem
\bibitem[{\citenamefont{Chen and Ma}(2017)}]{Chen:2016hgw}
\bibinfo{author}{\bibfnamefont{A.~P.} \bibnamefont{Chen}} \bibnamefont{and}
  \bibinfo{author}{\bibfnamefont{J.~P.} \bibnamefont{Ma}},
  \bibinfo{journal}{Phys. Lett.} \textbf{\bibinfo{volume}{B768}},
  \bibinfo{pages}{380} (\bibinfo{year}{2017}), \eprint{1610.08634}\relax
\mciteBstWouldAddEndPuncttrue
\mciteSetBstMidEndSepPunct{\mcitedefaultmidpunct}
{\mcitedefaultendpunct}{\mcitedefaultseppunct}\relax
\EndOfBibitem
\bibitem[{\citenamefont{Feige et~al.}(2017)\citenamefont{Feige, Kolodrubetz,
  Moult, and Stewart}}]{Feige:2017zci}
\bibinfo{author}{\bibfnamefont{I.}~\bibnamefont{Feige}},
  \bibinfo{author}{\bibfnamefont{D.~W.} \bibnamefont{Kolodrubetz}},
  \bibinfo{author}{\bibfnamefont{I.}~\bibnamefont{Moult}}, \bibnamefont{and}
  \bibinfo{author}{\bibfnamefont{I.~W.} \bibnamefont{Stewart}},
  \bibinfo{journal}{JHEP} \textbf{\bibinfo{volume}{11}}, \bibinfo{pages}{142}
  (\bibinfo{year}{2017}), \eprint{1703.03411}\relax
\mciteBstWouldAddEndPuncttrue
\mciteSetBstMidEndSepPunct{\mcitedefaultmidpunct}
{\mcitedefaultendpunct}{\mcitedefaultseppunct}\relax
\EndOfBibitem
\bibitem[{\citenamefont{Balitsky and Tarasov}(2017)}]{Balitsky:2017flc}
\bibinfo{author}{\bibfnamefont{I.}~\bibnamefont{Balitsky}} \bibnamefont{and}
  \bibinfo{author}{\bibfnamefont{A.}~\bibnamefont{Tarasov}},
  \bibinfo{journal}{JHEP} \textbf{\bibinfo{volume}{07}}, \bibinfo{pages}{095}
  (\bibinfo{year}{2017}), \eprint{1706.01415}\relax
\mciteBstWouldAddEndPuncttrue
\mciteSetBstMidEndSepPunct{\mcitedefaultmidpunct}
{\mcitedefaultendpunct}{\mcitedefaultseppunct}\relax
\EndOfBibitem
\bibitem[{\citenamefont{Balitsky and Tarasov}(2018)}]{Balitsky:2017gis}
\bibinfo{author}{\bibfnamefont{I.}~\bibnamefont{Balitsky}} \bibnamefont{and}
  \bibinfo{author}{\bibfnamefont{A.}~\bibnamefont{Tarasov}},
  \bibinfo{journal}{JHEP} \textbf{\bibinfo{volume}{05}}, \bibinfo{pages}{150}
  (\bibinfo{year}{2018}), \eprint{1712.09389}\relax
\mciteBstWouldAddEndPuncttrue
\mciteSetBstMidEndSepPunct{\mcitedefaultmidpunct}
{\mcitedefaultendpunct}{\mcitedefaultseppunct}\relax
\EndOfBibitem
\bibitem[{\citenamefont{Ebert et~al.}(2018)\citenamefont{Ebert, Moult, Stewart,
  Tackmann, Vita, and Zhu}}]{Ebert:2018lzn}
\bibinfo{author}{\bibfnamefont{M.~A.} \bibnamefont{Ebert}},
  \bibinfo{author}{\bibfnamefont{I.}~\bibnamefont{Moult}},
  \bibinfo{author}{\bibfnamefont{I.~W.} \bibnamefont{Stewart}},
  \bibinfo{author}{\bibfnamefont{F.~J.} \bibnamefont{Tackmann}},
  \bibinfo{author}{\bibfnamefont{G.}~\bibnamefont{Vita}}, \bibnamefont{and}
  \bibinfo{author}{\bibfnamefont{H.~X.} \bibnamefont{Zhu}},
  \bibinfo{journal}{JHEP} \textbf{\bibinfo{volume}{12}}, \bibinfo{pages}{084}
  (\bibinfo{year}{2018}), \eprint{1807.10764}\relax
\mciteBstWouldAddEndPuncttrue
\mciteSetBstMidEndSepPunct{\mcitedefaultmidpunct}
{\mcitedefaultendpunct}{\mcitedefaultseppunct}\relax
\EndOfBibitem
\bibitem[{\citenamefont{Ebert et~al.}(2019)\citenamefont{Ebert, Moult, Stewart,
  Tackmann, Vita, and Zhu}}]{Ebert:2018gsn}
\bibinfo{author}{\bibfnamefont{M.~A.} \bibnamefont{Ebert}},
  \bibinfo{author}{\bibfnamefont{I.}~\bibnamefont{Moult}},
  \bibinfo{author}{\bibfnamefont{I.~W.} \bibnamefont{Stewart}},
  \bibinfo{author}{\bibfnamefont{F.~J.} \bibnamefont{Tackmann}},
  \bibinfo{author}{\bibfnamefont{G.}~\bibnamefont{Vita}}, \bibnamefont{and}
  \bibinfo{author}{\bibfnamefont{H.~X.} \bibnamefont{Zhu}},
  \bibinfo{journal}{JHEP} \textbf{\bibinfo{volume}{04}}, \bibinfo{pages}{123}
  (\bibinfo{year}{2019}), \eprint{1812.08189}\relax
\mciteBstWouldAddEndPuncttrue
\mciteSetBstMidEndSepPunct{\mcitedefaultmidpunct}
{\mcitedefaultendpunct}{\mcitedefaultseppunct}\relax
\EndOfBibitem
\bibitem[{\citenamefont{Moult et~al.}(2019)\citenamefont{Moult, Stewart, and
  Vita}}]{Moult:2019mog}
\bibinfo{author}{\bibfnamefont{I.}~\bibnamefont{Moult}},
  \bibinfo{author}{\bibfnamefont{I.~W.} \bibnamefont{Stewart}},
  \bibnamefont{and} \bibinfo{author}{\bibfnamefont{G.}~\bibnamefont{Vita}}
  (\bibinfo{year}{2019}), \eprint{1905.07411}\relax
\mciteBstWouldAddEndPuncttrue
\mciteSetBstMidEndSepPunct{\mcitedefaultmidpunct}
{\mcitedefaultendpunct}{\mcitedefaultseppunct}\relax
\EndOfBibitem
\bibitem[{\citenamefont{Wandzura and Wilczek}(1977)}]{Wandzura:1977qf}
\bibinfo{author}{\bibfnamefont{S.}~\bibnamefont{Wandzura}} \bibnamefont{and}
  \bibinfo{author}{\bibfnamefont{F.}~\bibnamefont{Wilczek}},
  \bibinfo{journal}{Phys. Lett.} \textbf{\bibinfo{volume}{72B}},
  \bibinfo{pages}{195} (\bibinfo{year}{1977})\relax
\mciteBstWouldAddEndPuncttrue
\mciteSetBstMidEndSepPunct{\mcitedefaultmidpunct}
{\mcitedefaultendpunct}{\mcitedefaultseppunct}\relax
\EndOfBibitem
\bibitem[{\citenamefont{Barone and Ratcliffe}(2003)}]{Barone:2003fy}
\bibinfo{author}{\bibfnamefont{V.}~\bibnamefont{Barone}} \bibnamefont{and}
  \bibinfo{author}{\bibfnamefont{P.~G.} \bibnamefont{Ratcliffe}},
  \emph{\bibinfo{title}{{Transverse spin physics}}}
  (\bibinfo{year}{2003})\relax
\mciteBstWouldAddEndPuncttrue
\mciteSetBstMidEndSepPunct{\mcitedefaultmidpunct}
{\mcitedefaultendpunct}{\mcitedefaultseppunct}\relax
\EndOfBibitem
\bibitem[{\citenamefont{Bacchetta et~al.}(2004)\citenamefont{Bacchetta,
  D'Alesio, Diehl, and Miller}}]{Bacchetta:2004jz}
\bibinfo{author}{\bibfnamefont{A.}~\bibnamefont{Bacchetta}},
  \bibinfo{author}{\bibfnamefont{U.}~\bibnamefont{D'Alesio}},
  \bibinfo{author}{\bibfnamefont{M.}~\bibnamefont{Diehl}}, \bibnamefont{and}
  \bibinfo{author}{\bibfnamefont{C.~A.} \bibnamefont{Miller}},
  \bibinfo{journal}{Phys. Rev.} \textbf{\bibinfo{volume}{D70}},
  \bibinfo{pages}{117504} (\bibinfo{year}{2004}), \eprint{hep-ph/0410050}\relax
\mciteBstWouldAddEndPuncttrue
\mciteSetBstMidEndSepPunct{\mcitedefaultmidpunct}
{\mcitedefaultendpunct}{\mcitedefaultseppunct}\relax
\EndOfBibitem
\bibitem[{\citenamefont{Bacchetta et~al.}(2007)\citenamefont{Bacchetta, Diehl,
  Goeke, Metz, Mulders, and Schlegel}}]{Bacchetta:2006tn}
\bibinfo{author}{\bibfnamefont{A.}~\bibnamefont{Bacchetta}},
  \bibinfo{author}{\bibfnamefont{M.}~\bibnamefont{Diehl}},
  \bibinfo{author}{\bibfnamefont{K.}~\bibnamefont{Goeke}},
  \bibinfo{author}{\bibfnamefont{A.}~\bibnamefont{Metz}},
  \bibinfo{author}{\bibfnamefont{P.~J.} \bibnamefont{Mulders}},
  \bibnamefont{and} \bibinfo{author}{\bibfnamefont{M.}~\bibnamefont{Schlegel}},
  \bibinfo{journal}{JHEP} \textbf{\bibinfo{volume}{02}}, \bibinfo{pages}{093}
  (\bibinfo{year}{2007}), \eprint{hep-ph/0611265}\relax
\mciteBstWouldAddEndPuncttrue
\mciteSetBstMidEndSepPunct{\mcitedefaultmidpunct}
{\mcitedefaultendpunct}{\mcitedefaultseppunct}\relax
\EndOfBibitem
\bibitem[{\citenamefont{Boer et~al.}(2011)\citenamefont{Boer, Gamberg, Musch,
  and Prokudin}}]{Boer:2011xd}
\bibinfo{author}{\bibfnamefont{D.}~\bibnamefont{Boer}},
  \bibinfo{author}{\bibfnamefont{L.}~\bibnamefont{Gamberg}},
  \bibinfo{author}{\bibfnamefont{B.}~\bibnamefont{Musch}}, \bibnamefont{and}
  \bibinfo{author}{\bibfnamefont{A.}~\bibnamefont{Prokudin}},
  \bibinfo{journal}{JHEP} \textbf{\bibinfo{volume}{10}}, \bibinfo{pages}{021}
  (\bibinfo{year}{2011}), \eprint{1107.5294}\relax
\mciteBstWouldAddEndPuncttrue
\mciteSetBstMidEndSepPunct{\mcitedefaultmidpunct}
{\mcitedefaultendpunct}{\mcitedefaultseppunct}\relax
\EndOfBibitem
\bibitem[{\citenamefont{Cahn}(1978)}]{Cahn:1978se}
\bibinfo{author}{\bibfnamefont{R.~N.} \bibnamefont{Cahn}},
  \bibinfo{journal}{Phys. Lett.} \textbf{\bibinfo{volume}{78B}},
  \bibinfo{pages}{269} (\bibinfo{year}{1978})\relax
\mciteBstWouldAddEndPuncttrue
\mciteSetBstMidEndSepPunct{\mcitedefaultmidpunct}
{\mcitedefaultendpunct}{\mcitedefaultseppunct}\relax
\EndOfBibitem
\bibitem[{\citenamefont{Cahn}(1989)}]{Cahn:1989yf}
\bibinfo{author}{\bibfnamefont{R.~N.} \bibnamefont{Cahn}},
  \bibinfo{journal}{Phys. Rev.} \textbf{\bibinfo{volume}{D40}},
  \bibinfo{pages}{3107} (\bibinfo{year}{1989})\relax
\mciteBstWouldAddEndPuncttrue
\mciteSetBstMidEndSepPunct{\mcitedefaultmidpunct}
{\mcitedefaultendpunct}{\mcitedefaultseppunct}\relax
\EndOfBibitem
\bibitem[{\citenamefont{Arneodo et~al.}(1987)}]{Arneodo:1986cf}
\bibinfo{author}{\bibfnamefont{M.}~\bibnamefont{Arneodo}} \bibnamefont{et~al.}
  (\bibinfo{collaboration}{European Muon}), \bibinfo{journal}{Z. Phys.}
  \textbf{\bibinfo{volume}{C34}}, \bibinfo{pages}{277}
  (\bibinfo{year}{1987})\relax
\mciteBstWouldAddEndPuncttrue
\mciteSetBstMidEndSepPunct{\mcitedefaultmidpunct}
{\mcitedefaultendpunct}{\mcitedefaultseppunct}\relax
\EndOfBibitem
\bibitem[{\citenamefont{Adams et~al.}(1993)}]{Adams:1993hs}
\bibinfo{author}{\bibfnamefont{M.~R.} \bibnamefont{Adams}} \bibnamefont{et~al.}
  (\bibinfo{collaboration}{E665}), \bibinfo{journal}{Phys. Rev.}
  \textbf{\bibinfo{volume}{D48}}, \bibinfo{pages}{5057}
  (\bibinfo{year}{1993})\relax
\mciteBstWouldAddEndPuncttrue
\mciteSetBstMidEndSepPunct{\mcitedefaultmidpunct}
{\mcitedefaultendpunct}{\mcitedefaultseppunct}\relax
\EndOfBibitem
\bibitem[{\citenamefont{Breitweg et~al.}(2000)}]{Breitweg:2000qh}
\bibinfo{author}{\bibfnamefont{J.}~\bibnamefont{Breitweg}} \bibnamefont{et~al.}
  (\bibinfo{collaboration}{ZEUS}), \bibinfo{journal}{Phys. Lett.}
  \textbf{\bibinfo{volume}{B481}}, \bibinfo{pages}{199} (\bibinfo{year}{2000}),
  \eprint{hep-ex/0003017}\relax
\mciteBstWouldAddEndPuncttrue
\mciteSetBstMidEndSepPunct{\mcitedefaultmidpunct}
{\mcitedefaultendpunct}{\mcitedefaultseppunct}\relax
\EndOfBibitem
\bibitem[{\citenamefont{Chekanov et~al.}(2007)}]{Chekanov:2006gt}
\bibinfo{author}{\bibfnamefont{S.}~\bibnamefont{Chekanov}} \bibnamefont{et~al.}
  (\bibinfo{collaboration}{ZEUS}), \bibinfo{journal}{Eur. Phys. J.}
  \textbf{\bibinfo{volume}{C51}}, \bibinfo{pages}{289} (\bibinfo{year}{2007}),
  \eprint{hep-ex/0608053}\relax
\mciteBstWouldAddEndPuncttrue
\mciteSetBstMidEndSepPunct{\mcitedefaultmidpunct}
{\mcitedefaultendpunct}{\mcitedefaultseppunct}\relax
\EndOfBibitem
\bibitem[{\citenamefont{Airapetian et~al.}(2013)}]{Airapetian:2012yg}
\bibinfo{author}{\bibfnamefont{A.}~\bibnamefont{Airapetian}}
  \bibnamefont{et~al.} (\bibinfo{collaboration}{HERMES}),
  \bibinfo{journal}{Phys. Rev.} \textbf{\bibinfo{volume}{D87}},
  \bibinfo{pages}{012010} (\bibinfo{year}{2013}), \eprint{1204.4161}\relax
\mciteBstWouldAddEndPuncttrue
\mciteSetBstMidEndSepPunct{\mcitedefaultmidpunct}
{\mcitedefaultendpunct}{\mcitedefaultseppunct}\relax
\EndOfBibitem
\bibitem[{\citenamefont{Adolph et~al.}(2014)}]{Adolph:2014pwc}
\bibinfo{author}{\bibfnamefont{C.}~\bibnamefont{Adolph}} \bibnamefont{et~al.}
  (\bibinfo{collaboration}{COMPASS}), \bibinfo{journal}{Nucl. Phys.}
  \textbf{\bibinfo{volume}{B886}}, \bibinfo{pages}{1046}
  (\bibinfo{year}{2014}), \eprint{1401.6284}\relax
\mciteBstWouldAddEndPuncttrue
\mciteSetBstMidEndSepPunct{\mcitedefaultmidpunct}
{\mcitedefaultendpunct}{\mcitedefaultseppunct}\relax
\EndOfBibitem
\bibitem[{\citenamefont{Moretti}(2019)}]{Moretti:2019lkw}
\bibinfo{author}{\bibfnamefont{A.}~\bibnamefont{Moretti}}
  (\bibinfo{collaboration}{COMPASS}) (\bibinfo{year}{2019}),
  \eprint{1901.01773}\relax
\mciteBstWouldAddEndPuncttrue
\mciteSetBstMidEndSepPunct{\mcitedefaultmidpunct}
{\mcitedefaultendpunct}{\mcitedefaultseppunct}\relax
\EndOfBibitem
\bibitem[{\citenamefont{Anselmino et~al.}(2005)\citenamefont{Anselmino,
  Boglione, D'Alesio, Kotzinian, Murgia, and Prokudin}}]{Anselmino:2005nn}
\bibinfo{author}{\bibfnamefont{M.}~\bibnamefont{Anselmino}},
  \bibinfo{author}{\bibfnamefont{M.}~\bibnamefont{Boglione}},
  \bibinfo{author}{\bibfnamefont{U.}~\bibnamefont{D'Alesio}},
  \bibinfo{author}{\bibfnamefont{A.}~\bibnamefont{Kotzinian}},
  \bibinfo{author}{\bibfnamefont{F.}~\bibnamefont{Murgia}}, \bibnamefont{and}
  \bibinfo{author}{\bibfnamefont{A.}~\bibnamefont{Prokudin}},
  \bibinfo{journal}{Phys. Rev.} \textbf{\bibinfo{volume}{D71}},
  \bibinfo{pages}{074006} (\bibinfo{year}{2005}), \eprint{hep-ph/0501196}\relax
\mciteBstWouldAddEndPuncttrue
\mciteSetBstMidEndSepPunct{\mcitedefaultmidpunct}
{\mcitedefaultendpunct}{\mcitedefaultseppunct}\relax
\EndOfBibitem
\bibitem[{\citenamefont{Anselmino et~al.}(2007)\citenamefont{Anselmino,
  Boglione, Prokudin, and Turk}}]{Anselmino:2006rv}
\bibinfo{author}{\bibfnamefont{M.}~\bibnamefont{Anselmino}},
  \bibinfo{author}{\bibfnamefont{M.}~\bibnamefont{Boglione}},
  \bibinfo{author}{\bibfnamefont{A.}~\bibnamefont{Prokudin}}, \bibnamefont{and}
  \bibinfo{author}{\bibfnamefont{C.}~\bibnamefont{Turk}},
  \bibinfo{journal}{Eur. Phys. J.} \textbf{\bibinfo{volume}{A31}},
  \bibinfo{pages}{373} (\bibinfo{year}{2007}), \eprint{hep-ph/0606286}\relax
\mciteBstWouldAddEndPuncttrue
\mciteSetBstMidEndSepPunct{\mcitedefaultmidpunct}
{\mcitedefaultendpunct}{\mcitedefaultseppunct}\relax
\EndOfBibitem
\bibitem[{\citenamefont{Barone et~al.}(2015)\citenamefont{Barone, Boglione,
  Gonzalez~Hernandez, and Melis}}]{Barone:2015ksa}
\bibinfo{author}{\bibfnamefont{V.}~\bibnamefont{Barone}},
  \bibinfo{author}{\bibfnamefont{M.}~\bibnamefont{Boglione}},
  \bibinfo{author}{\bibfnamefont{J.~O.} \bibnamefont{Gonzalez~Hernandez}},
  \bibnamefont{and} \bibinfo{author}{\bibfnamefont{S.}~\bibnamefont{Melis}},
  \bibinfo{journal}{Phys. Rev.} \textbf{\bibinfo{volume}{D91}},
  \bibinfo{pages}{074019} (\bibinfo{year}{2015}), \eprint{1502.04214}\relax
\mciteBstWouldAddEndPuncttrue
\mciteSetBstMidEndSepPunct{\mcitedefaultmidpunct}
{\mcitedefaultendpunct}{\mcitedefaultseppunct}\relax
\EndOfBibitem
\bibitem[{\citenamefont{Mendez}(1978)}]{Mendez:1978zx}
\bibinfo{author}{\bibfnamefont{A.}~\bibnamefont{Mendez}},
  \bibinfo{journal}{Nucl. Phys.} \textbf{\bibinfo{volume}{B145}},
  \bibinfo{pages}{199} (\bibinfo{year}{1978})\relax
\mciteBstWouldAddEndPuncttrue
\mciteSetBstMidEndSepPunct{\mcitedefaultmidpunct}
{\mcitedefaultendpunct}{\mcitedefaultseppunct}\relax
\EndOfBibitem
\bibitem[{\citenamefont{Meng et~al.}(1996)\citenamefont{Meng, Olness, and
  Soper}}]{Meng:1995yn}
\bibinfo{author}{\bibfnamefont{R.}~\bibnamefont{Meng}},
  \bibinfo{author}{\bibfnamefont{F.~I.} \bibnamefont{Olness}},
  \bibnamefont{and} \bibinfo{author}{\bibfnamefont{D.~E.} \bibnamefont{Soper}},
  \bibinfo{journal}{Phys. Rev.} \textbf{\bibinfo{volume}{D54}},
  \bibinfo{pages}{1919} (\bibinfo{year}{1996}), \eprint{hep-ph/9511311}\relax
\mciteBstWouldAddEndPuncttrue
\mciteSetBstMidEndSepPunct{\mcitedefaultmidpunct}
{\mcitedefaultendpunct}{\mcitedefaultseppunct}\relax
\EndOfBibitem
\bibitem[{\citenamefont{Aybat and Rogers}(2011)}]{Aybat:2011zv}
\bibinfo{author}{\bibfnamefont{S.}~\bibnamefont{Aybat}} \bibnamefont{and}
  \bibinfo{author}{\bibfnamefont{T.~C.} \bibnamefont{Rogers}},
  \bibinfo{journal}{Phys. Rev.} \textbf{\bibinfo{volume}{D83}},
  \bibinfo{pages}{114042} (\bibinfo{year}{2011}), \eprint{1101.5057}\relax
\mciteBstWouldAddEndPuncttrue
\mciteSetBstMidEndSepPunct{\mcitedefaultmidpunct}
{\mcitedefaultendpunct}{\mcitedefaultseppunct}\relax
\EndOfBibitem
\bibitem[{\citenamefont{Echevarria et~al.}(2014)\citenamefont{Echevarria,
  Idilbi, and Scimemi}}]{Echevarria:2014rua}
\bibinfo{author}{\bibfnamefont{M.~G.} \bibnamefont{Echevarria}},
  \bibinfo{author}{\bibfnamefont{A.}~\bibnamefont{Idilbi}}, \bibnamefont{and}
  \bibinfo{author}{\bibfnamefont{I.}~\bibnamefont{Scimemi}},
  \bibinfo{journal}{Phys. Rev.} \textbf{\bibinfo{volume}{D90}},
  \bibinfo{pages}{014003} (\bibinfo{year}{2014}), \eprint{1402.0869}\relax
\mciteBstWouldAddEndPuncttrue
\mciteSetBstMidEndSepPunct{\mcitedefaultmidpunct}
{\mcitedefaultendpunct}{\mcitedefaultseppunct}\relax
\EndOfBibitem
\bibitem[{\citenamefont{Zhou et~al.}(2008)\citenamefont{Zhou, Yuan, and
  Liang}}]{Zhou:2008fb}
\bibinfo{author}{\bibfnamefont{J.}~\bibnamefont{Zhou}},
  \bibinfo{author}{\bibfnamefont{F.}~\bibnamefont{Yuan}}, \bibnamefont{and}
  \bibinfo{author}{\bibfnamefont{Z.-T.} \bibnamefont{Liang}},
  \bibinfo{journal}{Phys. Rev.} \textbf{\bibinfo{volume}{D78}},
  \bibinfo{pages}{114008} (\bibinfo{year}{2008}), \eprint{0808.3629}\relax
\mciteBstWouldAddEndPuncttrue
\mciteSetBstMidEndSepPunct{\mcitedefaultmidpunct}
{\mcitedefaultendpunct}{\mcitedefaultseppunct}\relax
\EndOfBibitem
\bibitem[{\citenamefont{Yuan and Zhou}(2009)}]{Yuan:2009dw}
\bibinfo{author}{\bibfnamefont{F.}~\bibnamefont{Yuan}} \bibnamefont{and}
  \bibinfo{author}{\bibfnamefont{J.}~\bibnamefont{Zhou}},
  \bibinfo{journal}{Phys. Rev. Lett.} \textbf{\bibinfo{volume}{103}},
  \bibinfo{pages}{052001} (\bibinfo{year}{2009}), \eprint{0903.4680}\relax
\mciteBstWouldAddEndPuncttrue
\mciteSetBstMidEndSepPunct{\mcitedefaultmidpunct}
{\mcitedefaultendpunct}{\mcitedefaultseppunct}\relax
\EndOfBibitem
\bibitem[{\citenamefont{Arnold et~al.}(2009)\citenamefont{Arnold, Metz, and
  Schlegel}}]{Arnold:2008kf}
\bibinfo{author}{\bibfnamefont{S.}~\bibnamefont{Arnold}},
  \bibinfo{author}{\bibfnamefont{A.}~\bibnamefont{Metz}}, \bibnamefont{and}
  \bibinfo{author}{\bibfnamefont{M.}~\bibnamefont{Schlegel}},
  \bibinfo{journal}{Phys. Rev.} \textbf{\bibinfo{volume}{D79}},
  \bibinfo{pages}{034005} (\bibinfo{year}{2009}), \eprint{0809.2262}\relax
\mciteBstWouldAddEndPuncttrue
\mciteSetBstMidEndSepPunct{\mcitedefaultmidpunct}
{\mcitedefaultendpunct}{\mcitedefaultseppunct}\relax
\EndOfBibitem
\bibitem[{\citenamefont{Lu and Schmidt}(2011{\natexlab{a}})}]{Lu:2011mz}
\bibinfo{author}{\bibfnamefont{Z.}~\bibnamefont{Lu}} \bibnamefont{and}
  \bibinfo{author}{\bibfnamefont{I.}~\bibnamefont{Schmidt}},
  \bibinfo{journal}{Phys. Rev.} \textbf{\bibinfo{volume}{D84}},
  \bibinfo{pages}{094002} (\bibinfo{year}{2011}{\natexlab{a}}),
  \eprint{1107.4693}\relax
\mciteBstWouldAddEndPuncttrue
\mciteSetBstMidEndSepPunct{\mcitedefaultmidpunct}
{\mcitedefaultendpunct}{\mcitedefaultseppunct}\relax
\EndOfBibitem
\bibitem[{\citenamefont{Lu and Schmidt}(2011{\natexlab{b}})}]{Lu:2011th}
\bibinfo{author}{\bibfnamefont{Z.}~\bibnamefont{Lu}} \bibnamefont{and}
  \bibinfo{author}{\bibfnamefont{I.}~\bibnamefont{Schmidt}},
  \bibinfo{journal}{Phys. Rev.} \textbf{\bibinfo{volume}{D84}},
  \bibinfo{pages}{114004} (\bibinfo{year}{2011}{\natexlab{b}}),
  \eprint{1109.3232}\relax
\mciteBstWouldAddEndPuncttrue
\mciteSetBstMidEndSepPunct{\mcitedefaultmidpunct}
{\mcitedefaultendpunct}{\mcitedefaultseppunct}\relax
\EndOfBibitem
\bibitem[{\citenamefont{Guanziroli et~al.}(1988)}]{Guanziroli:1987rp}
\bibinfo{author}{\bibfnamefont{M.}~\bibnamefont{Guanziroli}}
  \bibnamefont{et~al.} (\bibinfo{collaboration}{NA10}), \bibinfo{journal}{Z.
  Phys.} \textbf{\bibinfo{volume}{C37}}, \bibinfo{pages}{545}
  (\bibinfo{year}{1988})\relax
\mciteBstWouldAddEndPuncttrue
\mciteSetBstMidEndSepPunct{\mcitedefaultmidpunct}
{\mcitedefaultendpunct}{\mcitedefaultseppunct}\relax
\EndOfBibitem
\bibitem[{\citenamefont{Conway et~al.}(1989)}]{Conway:1989fs}
\bibinfo{author}{\bibfnamefont{J.~S.} \bibnamefont{Conway}}
  \bibnamefont{et~al.}, \bibinfo{journal}{Phys. Rev.}
  \textbf{\bibinfo{volume}{D39}}, \bibinfo{pages}{92}
  (\bibinfo{year}{1989})\relax
\mciteBstWouldAddEndPuncttrue
\mciteSetBstMidEndSepPunct{\mcitedefaultmidpunct}
{\mcitedefaultendpunct}{\mcitedefaultseppunct}\relax
\EndOfBibitem
\bibitem[{\citenamefont{Zhu et~al.}(2007)}]{Zhu:2006gx}
\bibinfo{author}{\bibfnamefont{L.~Y.} \bibnamefont{Zhu}} \bibnamefont{et~al.}
  (\bibinfo{collaboration}{NuSea}), \bibinfo{journal}{Phys. Rev. Lett.}
  \textbf{\bibinfo{volume}{99}}, \bibinfo{pages}{082301}
  (\bibinfo{year}{2007}), \eprint{hep-ex/0609005}\relax
\mciteBstWouldAddEndPuncttrue
\mciteSetBstMidEndSepPunct{\mcitedefaultmidpunct}
{\mcitedefaultendpunct}{\mcitedefaultseppunct}\relax
\EndOfBibitem
\bibitem[{\citenamefont{Zhu et~al.}(2009)}]{Zhu:2008sj}
\bibinfo{author}{\bibfnamefont{L.~Y.} \bibnamefont{Zhu}} \bibnamefont{et~al.}
  (\bibinfo{collaboration}{NuSea}), \bibinfo{journal}{Phys. Rev. Lett.}
  \textbf{\bibinfo{volume}{102}}, \bibinfo{pages}{182001}
  (\bibinfo{year}{2009}), \eprint{0811.4589}\relax
\mciteBstWouldAddEndPuncttrue
\mciteSetBstMidEndSepPunct{\mcitedefaultmidpunct}
{\mcitedefaultendpunct}{\mcitedefaultseppunct}\relax
\EndOfBibitem
\bibitem[{\citenamefont{Aaltonen et~al.}(2011)}]{Aaltonen:2011nr}
\bibinfo{author}{\bibfnamefont{T.}~\bibnamefont{Aaltonen}} \bibnamefont{et~al.}
  (\bibinfo{collaboration}{CDF}), \bibinfo{journal}{Phys. Rev. Lett.}
  \textbf{\bibinfo{volume}{106}}, \bibinfo{pages}{241801}
  (\bibinfo{year}{2011}), \eprint{1103.5699}\relax
\mciteBstWouldAddEndPuncttrue
\mciteSetBstMidEndSepPunct{\mcitedefaultmidpunct}
{\mcitedefaultendpunct}{\mcitedefaultseppunct}\relax
\EndOfBibitem
\bibitem[{\citenamefont{Khachatryan et~al.}(2015)}]{Khachatryan:2015paa}
\bibinfo{author}{\bibfnamefont{V.}~\bibnamefont{Khachatryan}}
  \bibnamefont{et~al.} (\bibinfo{collaboration}{CMS}), \bibinfo{journal}{Phys.
  Lett.} \textbf{\bibinfo{volume}{B750}}, \bibinfo{pages}{154}
  (\bibinfo{year}{2015}), \eprint{1504.03512}\relax
\mciteBstWouldAddEndPuncttrue
\mciteSetBstMidEndSepPunct{\mcitedefaultmidpunct}
{\mcitedefaultendpunct}{\mcitedefaultseppunct}\relax
\EndOfBibitem
\bibitem[{\citenamefont{Aad et~al.}(2016)}]{Aad:2016izn}
\bibinfo{author}{\bibfnamefont{G.}~\bibnamefont{Aad}} \bibnamefont{et~al.}
  (\bibinfo{collaboration}{ATLAS}), \bibinfo{journal}{JHEP}
  \textbf{\bibinfo{volume}{08}}, \bibinfo{pages}{159} (\bibinfo{year}{2016}),
  \eprint{1606.00689}\relax
\mciteBstWouldAddEndPuncttrue
\mciteSetBstMidEndSepPunct{\mcitedefaultmidpunct}
{\mcitedefaultendpunct}{\mcitedefaultseppunct}\relax
\EndOfBibitem
\bibitem[{\citenamefont{Barone et~al.}(2010)\citenamefont{Barone, Melis, and
  Prokudin}}]{Barone:2010gk}
\bibinfo{author}{\bibfnamefont{V.}~\bibnamefont{Barone}},
  \bibinfo{author}{\bibfnamefont{S.}~\bibnamefont{Melis}}, \bibnamefont{and}
  \bibinfo{author}{\bibfnamefont{A.}~\bibnamefont{Prokudin}},
  \bibinfo{journal}{Phys. Rev.} \textbf{\bibinfo{volume}{D82}},
  \bibinfo{pages}{114025} (\bibinfo{year}{2010}), \eprint{1009.3423}\relax
\mciteBstWouldAddEndPuncttrue
\mciteSetBstMidEndSepPunct{\mcitedefaultmidpunct}
{\mcitedefaultendpunct}{\mcitedefaultseppunct}\relax
\EndOfBibitem
\bibitem[{\citenamefont{Peng et~al.}(2016)\citenamefont{Peng, Chang, McClellan,
  and Teryaev}}]{Peng:2015spa}
\bibinfo{author}{\bibfnamefont{J.-C.} \bibnamefont{Peng}},
  \bibinfo{author}{\bibfnamefont{W.-C.} \bibnamefont{Chang}},
  \bibinfo{author}{\bibfnamefont{R.~E.} \bibnamefont{McClellan}},
  \bibnamefont{and} \bibinfo{author}{\bibfnamefont{O.}~\bibnamefont{Teryaev}},
  \bibinfo{journal}{Phys. Lett.} \textbf{\bibinfo{volume}{B758}},
  \bibinfo{pages}{384} (\bibinfo{year}{2016}), \eprint{1511.08932}\relax
\mciteBstWouldAddEndPuncttrue
\mciteSetBstMidEndSepPunct{\mcitedefaultmidpunct}
{\mcitedefaultendpunct}{\mcitedefaultseppunct}\relax
\EndOfBibitem
\bibitem[{\citenamefont{Lambertsen and Vogelsang}(2016)}]{Lambertsen:2016wgj}
\bibinfo{author}{\bibfnamefont{M.}~\bibnamefont{Lambertsen}} \bibnamefont{and}
  \bibinfo{author}{\bibfnamefont{W.}~\bibnamefont{Vogelsang}},
  \bibinfo{journal}{Phys. Rev.} \textbf{\bibinfo{volume}{D93}},
  \bibinfo{pages}{114013} (\bibinfo{year}{2016}), \eprint{1605.02625}\relax
\mciteBstWouldAddEndPuncttrue
\mciteSetBstMidEndSepPunct{\mcitedefaultmidpunct}
{\mcitedefaultendpunct}{\mcitedefaultseppunct}\relax
\EndOfBibitem
\bibitem[{\citenamefont{Motyka et~al.}(2017)\citenamefont{Motyka, Sadzikowski,
  and Stebel}}]{Motyka:2016lta}
\bibinfo{author}{\bibfnamefont{L.}~\bibnamefont{Motyka}},
  \bibinfo{author}{\bibfnamefont{M.}~\bibnamefont{Sadzikowski}},
  \bibnamefont{and} \bibinfo{author}{\bibfnamefont{T.}~\bibnamefont{Stebel}},
  \bibinfo{journal}{Phys. Rev.} \textbf{\bibinfo{volume}{D95}},
  \bibinfo{pages}{114025} (\bibinfo{year}{2017}), \eprint{1609.04300}\relax
\mciteBstWouldAddEndPuncttrue
\mciteSetBstMidEndSepPunct{\mcitedefaultmidpunct}
{\mcitedefaultendpunct}{\mcitedefaultseppunct}\relax
\EndOfBibitem
\bibitem[{\citenamefont{Peng et~al.}(2019)\citenamefont{Peng, Boer, Chang,
  McClellan, and Teryaev}}]{Peng:2018tty}
\bibinfo{author}{\bibfnamefont{J.-C.} \bibnamefont{Peng}},
  \bibinfo{author}{\bibfnamefont{D.}~\bibnamefont{Boer}},
  \bibinfo{author}{\bibfnamefont{W.-C.} \bibnamefont{Chang}},
  \bibinfo{author}{\bibfnamefont{R.~E.} \bibnamefont{McClellan}},
  \bibnamefont{and} \bibinfo{author}{\bibfnamefont{O.}~\bibnamefont{Teryaev}},
  \bibinfo{journal}{Phys. Lett.} \textbf{\bibinfo{volume}{B789}},
  \bibinfo{pages}{356} (\bibinfo{year}{2019}), \eprint{1808.04398}\relax
\mciteBstWouldAddEndPuncttrue
\mciteSetBstMidEndSepPunct{\mcitedefaultmidpunct}
{\mcitedefaultendpunct}{\mcitedefaultseppunct}\relax
\EndOfBibitem
\bibitem[{\citenamefont{Collins and Soper}(1977)}]{Collins:1977iv}
\bibinfo{author}{\bibfnamefont{J.~C.} \bibnamefont{Collins}} \bibnamefont{and}
  \bibinfo{author}{\bibfnamefont{D.~E.} \bibnamefont{Soper}},
  \bibinfo{journal}{Phys. Rev.} \textbf{\bibinfo{volume}{D16}},
  \bibinfo{pages}{2219} (\bibinfo{year}{1977})\relax
\mciteBstWouldAddEndPuncttrue
\mciteSetBstMidEndSepPunct{\mcitedefaultmidpunct}
{\mcitedefaultendpunct}{\mcitedefaultseppunct}\relax
\EndOfBibitem
\bibitem[{\citenamefont{Lam and Tung}(1978)}]{Lam:1978pu}
\bibinfo{author}{\bibfnamefont{C.~S.} \bibnamefont{Lam}} \bibnamefont{and}
  \bibinfo{author}{\bibfnamefont{W.-K.} \bibnamefont{Tung}},
  \bibinfo{journal}{Phys. Rev.} \textbf{\bibinfo{volume}{D18}},
  \bibinfo{pages}{2447} (\bibinfo{year}{1978})\relax
\mciteBstWouldAddEndPuncttrue
\mciteSetBstMidEndSepPunct{\mcitedefaultmidpunct}
{\mcitedefaultendpunct}{\mcitedefaultseppunct}\relax
\EndOfBibitem
\bibitem[{\citenamefont{Gauld et~al.}(2017)\citenamefont{Gauld,
  Gehrmann-De~Ridder, Gehrmann, Glover, and Huss}}]{Gauld:2017tww}
\bibinfo{author}{\bibfnamefont{R.}~\bibnamefont{Gauld}},
  \bibinfo{author}{\bibfnamefont{A.}~\bibnamefont{Gehrmann-De~Ridder}},
  \bibinfo{author}{\bibfnamefont{T.}~\bibnamefont{Gehrmann}},
  \bibinfo{author}{\bibfnamefont{E.~W.~N.} \bibnamefont{Glover}},
  \bibnamefont{and} \bibinfo{author}{\bibfnamefont{A.}~\bibnamefont{Huss}},
  \bibinfo{journal}{JHEP} \textbf{\bibinfo{volume}{11}}, \bibinfo{pages}{003}
  (\bibinfo{year}{2017}), \eprint{1708.00008}\relax
\mciteBstWouldAddEndPuncttrue
\mciteSetBstMidEndSepPunct{\mcitedefaultmidpunct}
{\mcitedefaultendpunct}{\mcitedefaultseppunct}\relax
\EndOfBibitem
\bibitem[{\citenamefont{Kanazawa et~al.}(2016)\citenamefont{Kanazawa, Koike,
  Metz, Pitonyak, and Schlegel}}]{Kanazawa:2015ajw}
\bibinfo{author}{\bibfnamefont{K.}~\bibnamefont{Kanazawa}},
  \bibinfo{author}{\bibfnamefont{Y.}~\bibnamefont{Koike}},
  \bibinfo{author}{\bibfnamefont{A.}~\bibnamefont{Metz}},
  \bibinfo{author}{\bibfnamefont{D.}~\bibnamefont{Pitonyak}}, \bibnamefont{and}
  \bibinfo{author}{\bibfnamefont{M.}~\bibnamefont{Schlegel}},
  \bibinfo{journal}{Phys. Rev.} \textbf{\bibinfo{volume}{D93}},
  \bibinfo{pages}{054024} (\bibinfo{year}{2016}), \eprint{1512.07233}\relax
\mciteBstWouldAddEndPuncttrue
\mciteSetBstMidEndSepPunct{\mcitedefaultmidpunct}
{\mcitedefaultendpunct}{\mcitedefaultseppunct}\relax
\EndOfBibitem
\bibitem[{\citenamefont{Boer}(2009)}]{Boer:2008fr}
\bibinfo{author}{\bibfnamefont{D.}~\bibnamefont{Boer}}, \bibinfo{journal}{Nucl.
  Phys.} \textbf{\bibinfo{volume}{B806}}, \bibinfo{pages}{23}
  (\bibinfo{year}{2009}), \eprint{0804.2408}\relax
\mciteBstWouldAddEndPuncttrue
\mciteSetBstMidEndSepPunct{\mcitedefaultmidpunct}
{\mcitedefaultendpunct}{\mcitedefaultseppunct}\relax
\EndOfBibitem
\end{mcitethebibliography}

\end{document}